%% file: main.tex
\documentclass[10pt,conference]{IEEEtran}
\usepackage{cite}
\usepackage{amsmath,amssymb}
\usepackage{graphicx}
\usepackage{xcolor}
\usepackage[hyphens]{url}
\usepackage[hidelinks]{hyperref}
\usepackage{booktabs}
\usepackage{multirow}
\usepackage{xspace}
\usepackage[skins, breakable]{tcolorbox}

\graphicspath{{figures/}}
\DeclareRobustCommand{\HhRay}{H\textsuperscript{2}-Ray\xspace}
\newcommand{\HhRayPlain}{H2-Ray}
\newcommand{\NumEvalScenes}{8}
\newcommand{\EvalGeoSpeedup}{3.77}
\newcommand{\EvalMaxSpeedup}{9.26}
\newcommand{\EvalMinSpeedup}{1.89}
\newcommand{\RaySplatsAvgRuntimeMs}{514.59}
\newcommand{\HhRayAvgRuntimeMs}{131.65}
\newcommand{\DrJohnsonSpeedup}{5.58}
\newcommand{\PlayroomSpeedup}{5.36}
\newcommand{\AFiveKRTTFLOPS}{54.2}
\newcommand{\HhRayScaledRTTFLOPS}{54.315}
\newcommand{\LBSBVHMinSpeedup}{1.08}
\newcommand{\LBSBVHMaxSpeedup}{1.31}
\newcommand{\LBSBVHAvgSpeedup}{1.18}
\newcommand{\LBSBVHOneGridAvg}{9,890}
\newcommand{\LBSBVHTwentySevenGridAvg}{8,380}
\newcommand{\LBSBVHDRAMMin}{5.9}
\newcommand{\LBSBVHDRAMMax}{31.3}
\newcommand{\LBSBVHDRAMAvg}{18.9}
\newcommand{\QualityAvgPSNRDelta}{-0.034}
\newcommand{\QualityAvgPSNRDrop}{0.034}
\newcommand{\RCPRTwoHopAvgLoss}{0.06}
\newcommand{\RCPRTwoHopMaxLoss}{0.13}
\newcommand{\UIESharedFPUnits}{31}
\newcommand{\UIETotalFPUnits}{78}
\newcommand{\UIESharedFPPct}{40}
\newcommand{\UIEGaussianFPUnits}{32}
\newcommand{\UIEGaussianFPPct}{41}
\newcommand{\UIETriangleFPUnits}{15}
\newcommand{\UIETriangleFPPct}{19}

\newtcolorbox{takeawaybox}{
    enhanced,
    breakable,
    colback=gray!10,
    colframe=gray!50,
    arc=2pt,
    boxrule=0.5pt,
    left=5pt, right=5pt,
    top=4pt, bottom=4pt,
    boxsep=0pt,
}

\title{A Hybrid and Hierarchical Ray Tracing Method for Efficient 3D Gaussian Rendering}

\author{
  \IEEEauthorblockN{
    Tiancheng Zhao\textsuperscript{1,2},
    Sixu Li\textsuperscript{2},
    Yuzhou Chen\textsuperscript{3},
    Hengyu Ding\textsuperscript{2},
    Zhiming Pan\textsuperscript{4},
    Chaojian Li\textsuperscript{1}
  }

  \IEEEauthorblockA{
    \textsuperscript{1}Hong Kong University of Science and Technology \\
    \textsuperscript{2}Georgia Institute of Technology \\
    \textsuperscript{3}Shanghai Jiao Tong University \\
    \textsuperscript{4}Peking University
  }
}

\hypersetup{
  pdftitle={A Hybrid and Hierarchical Ray Tracing Method for Efficient 3D Gaussian Rendering},
  pdfauthor={Tiancheng Zhao, Sixu Li, Yuzhou Chen, Hengyu Ding, Zhiming Pan, Chaojian Li},
  pdfkeywords={Neural Rendering, 3D Gaussian Splatting, Ray Tracing, GPU Architecture, Hardware Acceleration}
}

\begin{document}

\maketitle
\thispagestyle{plain}
\pagestyle{plain}

\begin{abstract}

3D Gaussian Splatting (3DGS) is a powerful representation for real-time neural reconstruction and rendering.
While existing 3DGS systems rely on rasterization for efficiency, recent work explores ray tracing to enable secondary lighting, highly distorted cameras, and depth-of-field effects.
However, compared with rasterization, ray-traced 3DGS introduces three major challenges:
(1) the highly non-uniform spatial distribution of massive numbers of Gaussian primitives creates severe load imbalance;
(2) neighboring pixels often generate rays with highly similar traversal paths, leading to substantial redundant traversal overhead; and
(3) supporting both Gaussian and triangle primitives yields heterogeneous intersection workloads that are not well matched to existing GPU architectures.
We present \HhRay, a hybrid and hierarchical ray tracing architecture for efficient 3D Gaussian rendering. \HhRay introduces three techniques.
Load-Balanced Spatial BVH (LB-SBVH) partitions the scene into uniform grids and uses density-aware representations to convert an imbalanced global workload into balanced local traversal units.
Ray-Cluster Path Reuse (RCPR) lets a proxy ray share traversal state with neighboring rays to amortize redundant traversal.
The Unified Intersection Engine (UIE) supports mixed Gaussian and triangle intersections through a mode-multiplexed datapath that shares \UIESharedFPPct\% of its floating-point units across both primitive types.
Across \NumEvalScenes{} scenes, \HhRay achieves a geometric-mean speedup of \EvalGeoSpeedup$\times$ and a maximum speedup of \EvalMaxSpeedup$\times$ over RaySplats on the NVIDIA RTX A5000, combining a cycle-accurate model of the modified RT Core with measured CUDA timings.

\end{abstract}


\input{sections/01-Intro}
\input{sections/02-Preliminary}
\input{sections/03-Algorithm}
\input{sections/04-Codesign}
\input{sections/06-Evaluation}
\input{sections/07-RelatedWorks}

\section{Conclusion}

\HhRay combines LB-SBVH scene partitioning, RCPR traversal reuse, and unified Gaussian--triangle intersection to address spatial imbalance, redundant traversal, and heterogeneous primitives in ray-traced 3DGS. Across \NumEvalScenes{} scenes, it achieves a \EvalGeoSpeedup$\times$ geometric-mean speedup over RaySplats while maintaining small average PSNR differences across the matched views.

\input{main.bbl}
\end{document}

%% file: sections/01-Intro.tex
\section{Introduction}

\begin{figure}[t]
    \centering
    \includegraphics[width=\columnwidth]{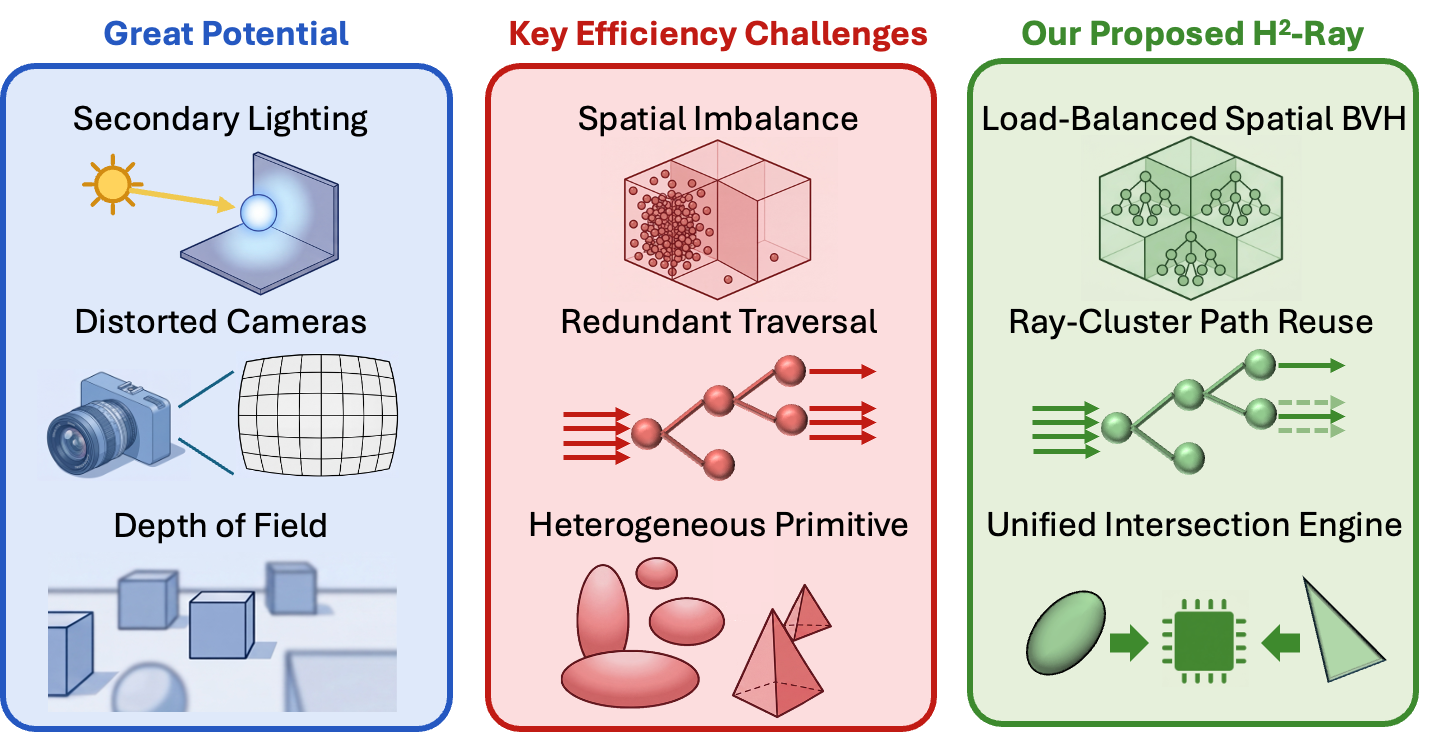}
    \caption{
    Applying ray tracing to 3D Gaussian rendering enables advanced optical effects but introduces three key challenges: spatial imbalance, redundant traversal, and heterogeneous primitive intersections. \HhRay addresses these challenges with a Load-Balanced Spatial BVH, Ray-Cluster Path Reuse, and a Unified Intersection Engine, respectively.
    }
    \label{fig:intro-teaser}
\end{figure}

3D Gaussian Splatting (3DGS) has become a compelling scene representation for neural reconstruction and rendering, delivering high visual quality at real-time rates~\cite{kerbl3Dgaussians}.
Its anisotropic primitives and rasterization-friendly rendering capture fine geometric and appearance details at interactive rates.
However, the rasterization-centric design that makes 3DGS efficient also limits its ability to model real-world optical effects.
In particular, rasterization-based Gaussian rendering struggles to support (1) secondary-lighting effects such as reflections and refractions needed for mirror-like and transparent materials~\cite{pbrt4e}, (2) highly distorted fisheye cameras used in autonomous-driving surround-view systems~\cite{woodscape}, (3) depth-of-field effects arising from finite-aperture camera optics~\cite{distributedRayTracing}, and other rendering modes that benefit from per-ray visibility reasoning.
As summarized in Figure~\ref{fig:intro-teaser}, ray tracing broadens the rendering envelope of 3DGS, but making this flexibility practical requires architectural support beyond both triangle-centric ray tracing hardware and rasterization-based 3DGS pipelines.

\begin{figure*}[t]
    \centering

    \includegraphics[width=0.243\textwidth]{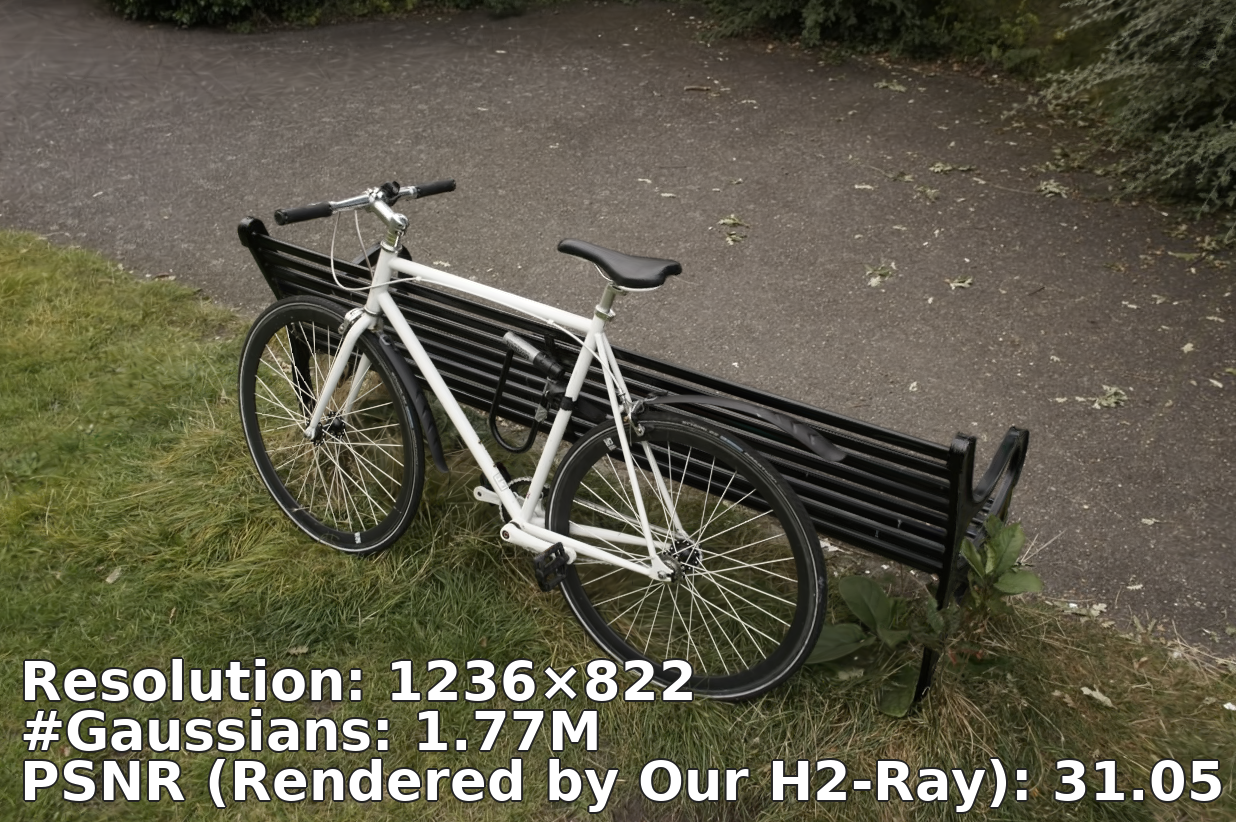}\hfill
    \includegraphics[width=0.243\textwidth]{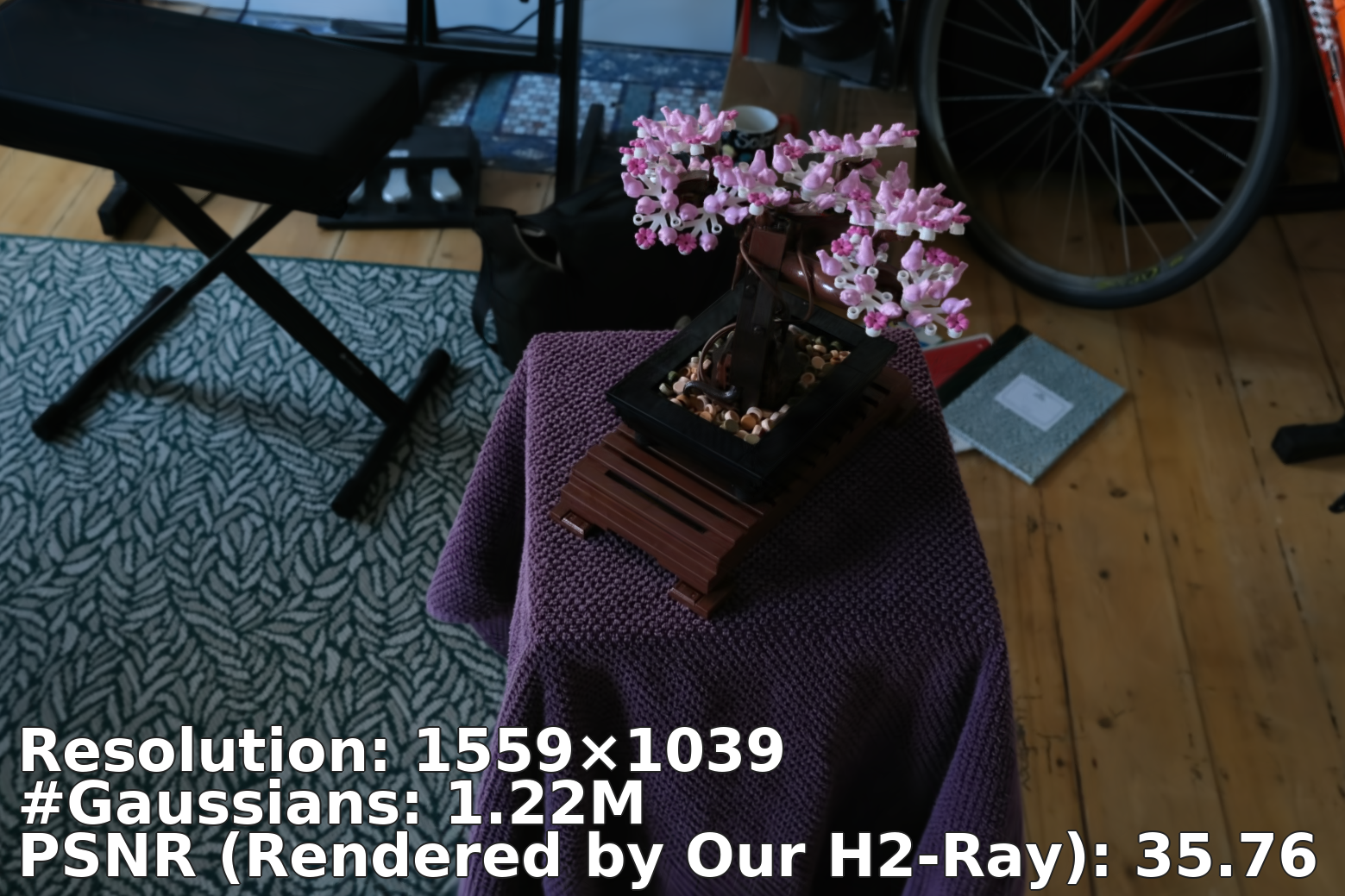}\hfill
    \includegraphics[width=0.243\textwidth]{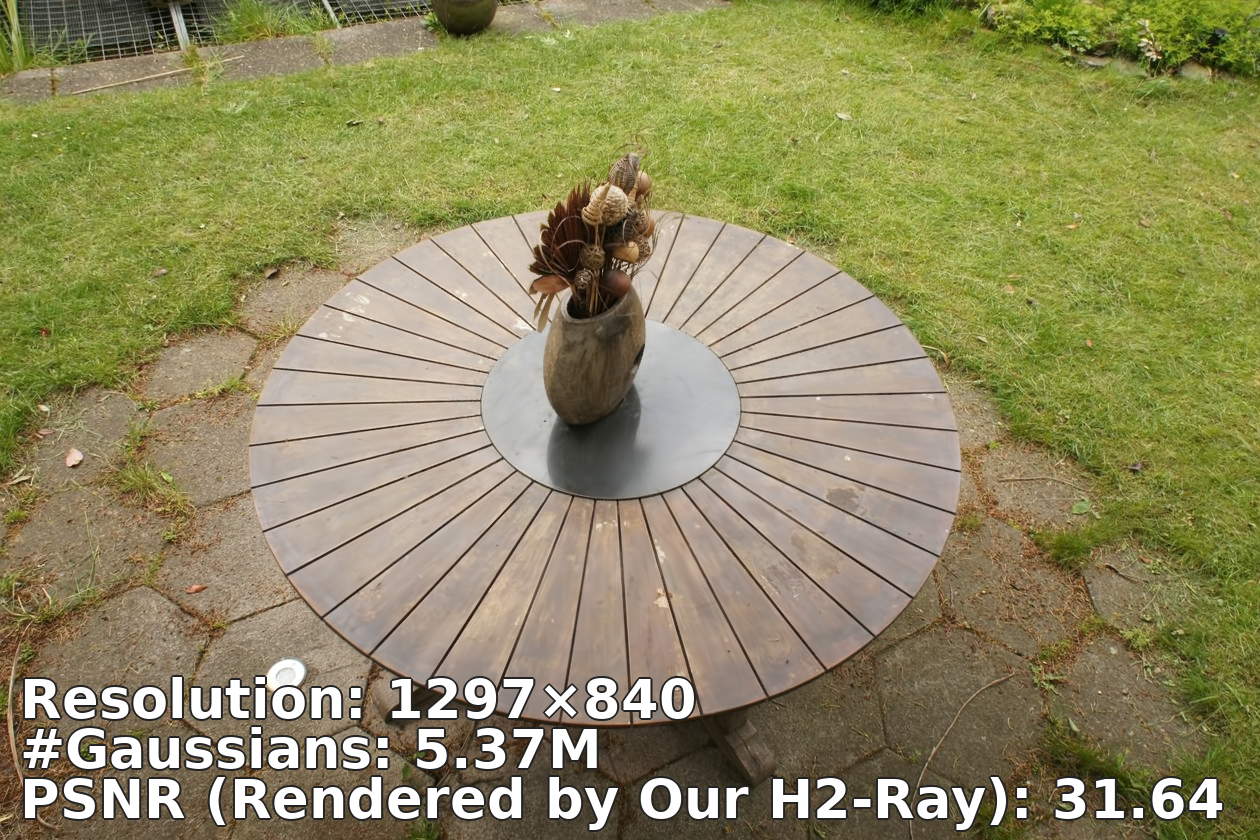}\hfill
    \includegraphics[width=0.243\textwidth]{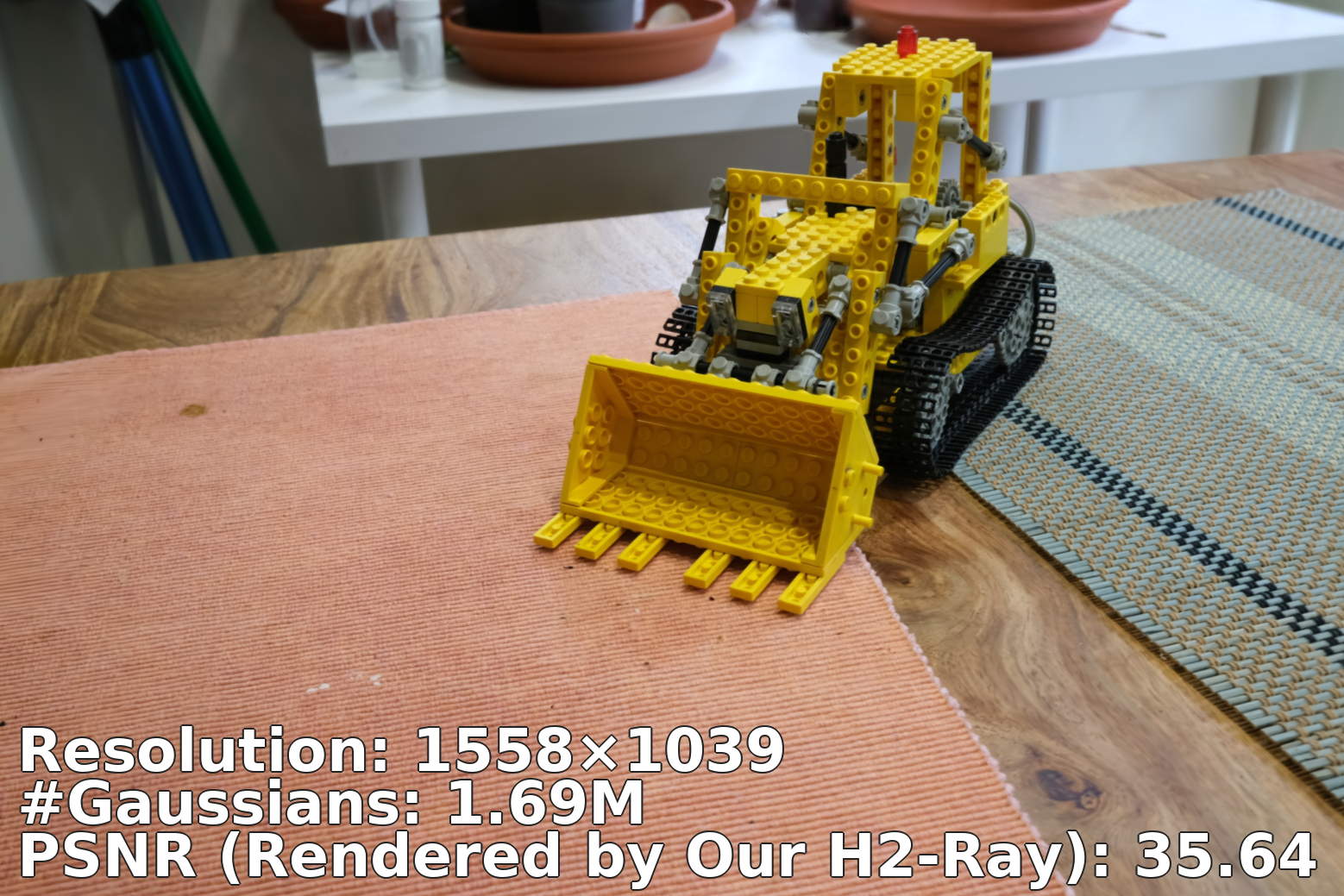}

    \vspace{2pt}

    \includegraphics[width=0.243\textwidth]{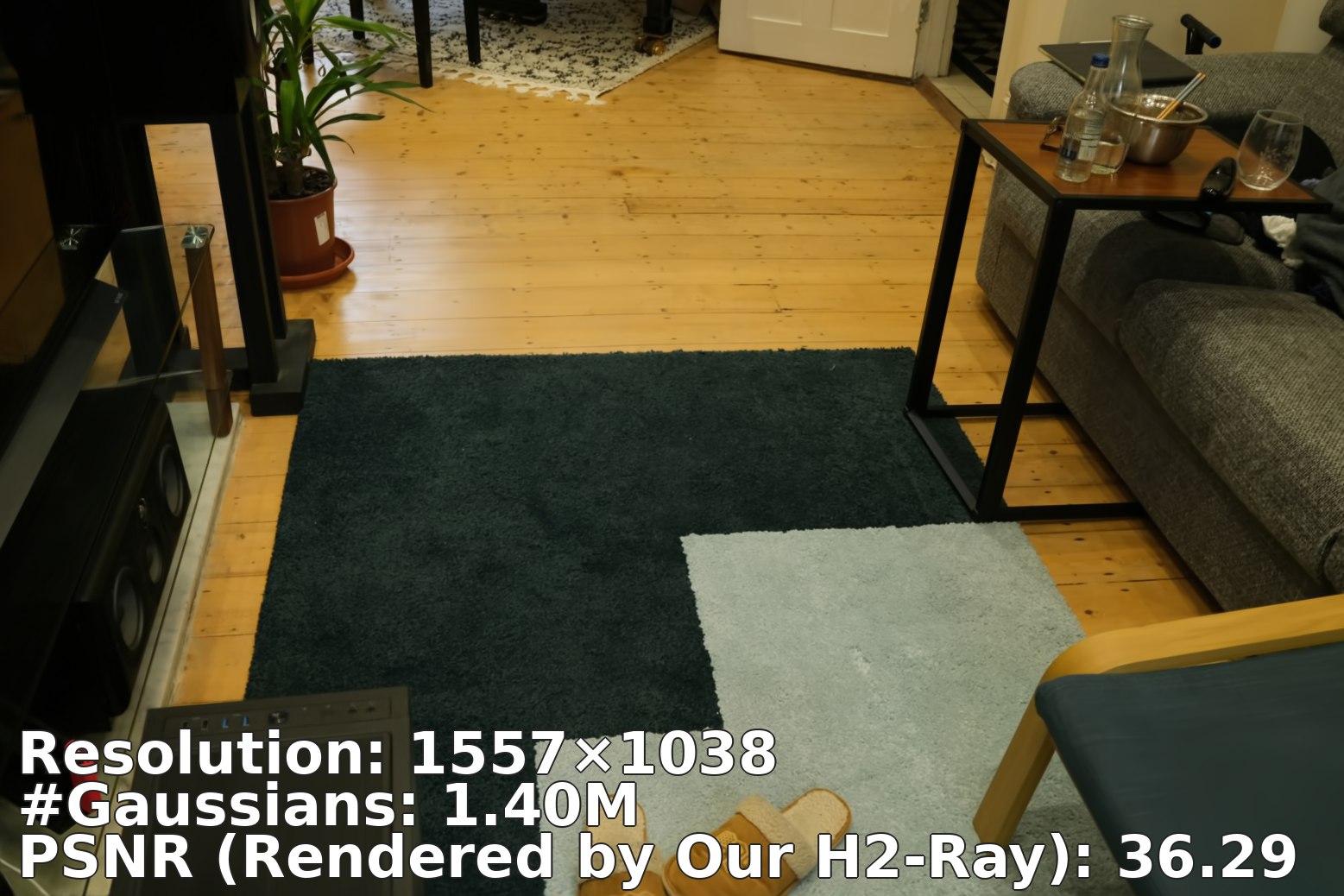}\hfill
    \includegraphics[width=0.243\textwidth]{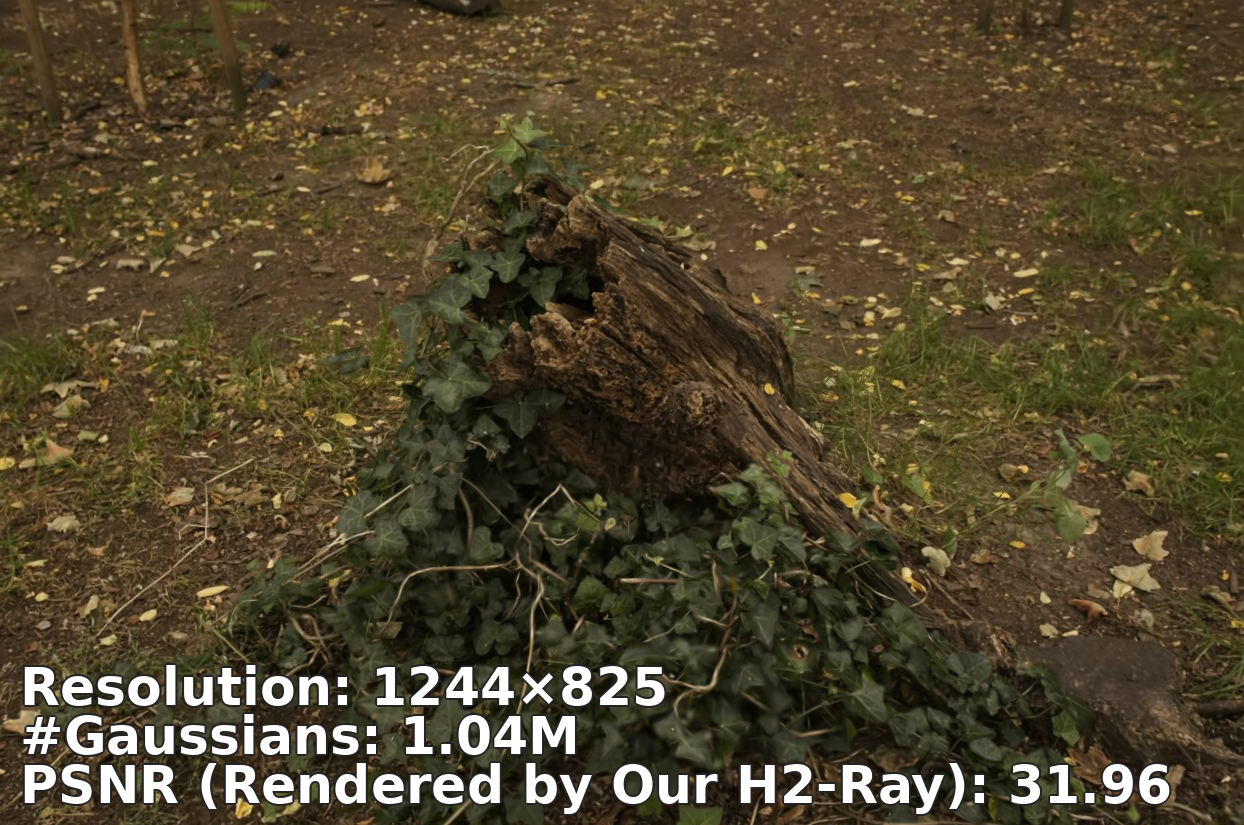}\hfill
    \includegraphics[width=0.243\textwidth]{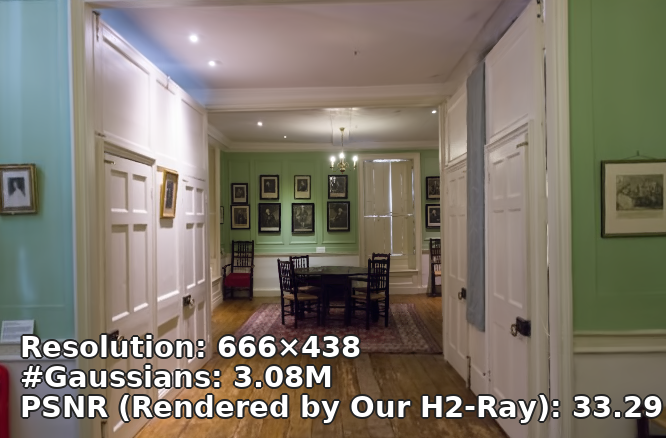}\hfill
    \includegraphics[width=0.243\textwidth]{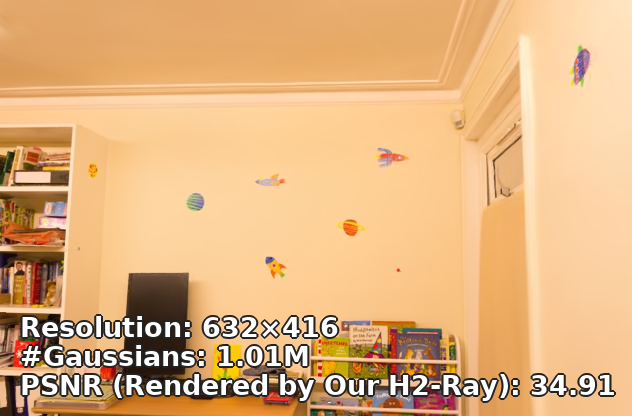}

    \caption{Preview of \NumEvalScenes{} representative scenes from Mip-NeRF 360~\cite{mipnerf360} and Deep Blending~\cite{deepblending} rendered using \HhRay. Each PSNR annotation is computed for the displayed representative view against its ground-truth image, rather than averaged across views.}
    \label{fig:scene-preview-8}
\end{figure*}

This gap stems from 3D Gaussian ray-tracing workloads, which differ from rasterization and triangle-based ray tracing.
\underline{First}, Gaussian primitives are highly non-uniform across space, so a monolithic acceleration structure must simultaneously serve extremely dense and extremely sparse regions, leading to poor traversal balance and inefficient parallelism.
In rasterization, imbalance mainly arises from the number of projected fragments in 2D screen space. In ray tracing, which operates in 3D, secondary-ray effects (e.g., reflections and refractions) further amplify spatial non-uniformity, resulting in more severe load imbalance and irregular execution.
\underline{Second}, neighboring pixels often generate rays that traverse highly similar BVH path prefixes, so processing these rays independently causes substantial redundant traversal work.
This redundancy is most pronounced at high resolutions, where dense pixel sampling induces strong spatial coherence among rays and deep BVH hierarchies lead to long shared traversal prefixes.
\underline{Third}, advanced effects increasingly mix ray--Gaussian and ray--triangle intersections, creating a heterogeneous workload mismatched to fixed-function specialization around one dominant primitive type.
Modern GPU architectures provide dedicated hardware support for ray--triangle intersections (e.g., via RT Cores), while ray--Gaussian intersections are typically handled by general-purpose compute units, leading to an imbalance in execution efficiency.
This heterogeneity limits hardware efficiency and motivates unified support for both primitives.

Recent efforts have brought ray tracing to 3D Gaussian rendering.
3DGRT~\cite{moenneLoccoz3DGRT}, 3DGUT~\cite{wu3DGUT}, and RaySplats~\cite{byrskiRaySplats} demonstrate ray-traced 3DGS with different rendering formulations, and GRTX~\cite{GRTX} adds hardware support for Gaussian BVH traversal.
However, these works target rendering formulations or single-ray traversal, leaving spatial imbalance, redundant traversal across neighboring rays, and heterogeneous intersection largely unexplored.

To address these challenges, we present \HhRay, a \textbf{hybrid} and \textbf{hierarchical} ray tracing architecture.
\HhRay is \underline{\emph{hybrid}} in two senses: it combines rasterization with ray tracing for fast region-of-interest culling, and it supports mixed ray--Gaussian and ray--triangle intersection workloads.
It is \underline{\emph{hierarchical}} because it addresses the workload at three levels: scene-level spatial partitioning, ray-level traversal reuse, and intersection-level execution adaptation.
Rather than replacing modern graphics hardware, \HhRay adopts a practical co-design approach, adding targeted support at key points in the GPU rendering pipeline. Our contributions are as follows.

\begin{itemize}

    \item We propose Load-Balanced Spatial BVH (\textbf{LB-SBVH}), a density-aware spatial partitioning scheme that converts globally skewed Gaussian distributions into balanced traversal regions, achieving a microarchitectural traversal speedup of up to $\LBSBVHMaxSpeedup\times$ over a monolithic BVH.

    \item We propose Ray-Cluster Path Reuse (\textbf{RCPR}), which shares traversal across neighboring rays to reduce redundant BVH operations. In matched multi-view evaluation, \HhRay decreases average PSNR by only \QualityAvgPSNRDrop\,dB relative to RaySplats.

    \item We design a Unified Intersection Engine (\textbf{UIE}) that supports both Gaussian and triangle primitives within a single reconfigurable datapath, sharing \UIESharedFPUnits{} of its \UIETotalFPUnits{} floating-point units (approximately \UIESharedFPPct\%) across primitive types, with \UIETriangleFPUnits{} units (approximately \UIETriangleFPPct\%) dedicated to triangles.

\end{itemize}

We evaluate \HhRay on \NumEvalScenes{} scenes from Mip-NeRF 360~\cite{mipnerf360} and Deep Blending~\cite{deepblending} (Figure~\ref{fig:scene-preview-8}), combining cycle-level modeling of the modified RT Core with measured RTX A5000 timings for all CUDA work.
\HhRay achieves a geometric-mean speedup of \EvalGeoSpeedup$\times$ (up to \EvalMaxSpeedup$\times$) over RaySplats on the A5000 while enabling rendering effects beyond rasterization-based pipelines.

%% file: sections/02-Preliminary.tex
\section{Background}

\subsection{3D Gaussian Splatting}
3DGS represents a scene as a set of anisotropic Gaussian primitives, where each primitive stores a mean position, covariance, opacity, and view-dependent appearance~\cite{kerbl3Dgaussians}. A Gaussian primitive can be written as
\begin{equation}
\label{eq:gaussian}
G_i(\mathbf{x}) = \alpha_i \exp \left( -\frac{1}{2} (\mathbf{x}-\boldsymbol{\mu}_i)^T \boldsymbol{\Sigma}_i^{-1} (\mathbf{x}-\boldsymbol{\mu}_i) \right),
\end{equation}
where $\boldsymbol{\mu}_i$ and $\boldsymbol{\Sigma}_i$ describe the primitive's spatial support, and $\alpha_i$ controls its contribution strength. In practice, a reconstructed scene contains a very large number of such Gaussians, and their spatial distribution is often highly non-uniform, as quantified later in Table~\ref{tab:motivation-grid-stats} and Figure~\ref{fig:motivation-imbalance}.

Rasterization-based 3DGS projects these 3D Gaussians onto the image plane, bins them into screen-space tiles, and blends their contributions in front-to-back order. The rendered color of a pixel can be written in the standard alpha-compositing form
\begin{equation}
C(\mathbf{r}) = \sum_{i} T_i \, \hat{\alpha}_i \, \mathbf{c}_i,
\qquad
T_i = \prod_{j<i} (1-\hat{\alpha}_j),
\end{equation}
where $\hat{\alpha}_i$ is the projected opacity of the $i$-th Gaussian along ray $\mathbf{r}$, $\mathbf{c}_i$ is its color, and $T_i$ is the accumulated transmittance before it. This rasterization pipeline is highly efficient for standard pinhole-camera novel-view synthesis, but it is less natural for rendering modes that require arbitrary-ray queries, secondary visibility, or tight integration with conventional ray-traced geometry.

\subsection{Ray Tracing for 3DGS Rendering}
Recent work revisits 3DGS rendering from a ray-based viewpoint.
3DGRT~\cite{moenneLoccoz3DGRT} builds a BVH over Gaussian primitives, encapsulates each with a stretched bounding mesh, and uses hardware ray--triangle intersection.
This design improves flexibility, but traversal and ordering remain proxy-based: primitives are gathered through bounding-mesh hits, and integration follows proxy intersections rather than exact sample locations along the ray.
3DGUT~\cite{wu3DGUT} instead preserves splatting for primary rendering while extending 3DGS to nonlinear cameras through Unscented-Transform sigma points.
3DGUT further aligns its rendering formulation with 3DGRT so that primary rays can still be splatted efficiently while secondary rays are traced in the same representation.
This bridges rasterization efficiency and ray-tracing flexibility, but primary rays still use a sigma-point approximation rather than direct ray--Gaussian intersection, while secondary rays follow the ray-traced representation.
These designs emphasize rendering flexibility and camera-model support, whereas our architectural study focuses on the execution costs exposed by direct Gaussian traversal.

RaySplats~\cite{byrskiRaySplats} moves further toward a direct ray-based formulation. Instead of using bounding polytopes as the primitive representation, RaySplats explicitly computes intersections between rays and Gaussian confidence ellipsoids, avoiding polytope approximation and remaining applicable to flat Gaussians.
Importantly, this ellipsoid is not a different primitive from \eqref{eq:gaussian}; it is an iso-surface induced by the same Gaussian after choosing a fixed density threshold. Specifically, setting $G_i(\mathbf{x})=\rho$ in \eqref{eq:gaussian} yields
\begin{equation}
(\mathbf{x}-\boldsymbol{\mu})^T \boldsymbol{\Sigma}^{-1} (\mathbf{x}-\boldsymbol{\mu}) = \tau,
\end{equation}
where $\tau = -2 \ln (\rho / \alpha)$ is a threshold-dependent constant. For a ray $\mathbf{r}(t)=\mathbf{o}+t\mathbf{d}$, substituting $\mathbf{x}=\mathbf{o}+t\mathbf{d}$ into this iso-surface equation yields a quadratic
\begin{equation}
a t^2 + b t + c = 0,
\end{equation}
where
\[
\begin{aligned}
a &= \mathbf{d}^T \boldsymbol{\Sigma}^{-1}\mathbf{d}, \qquad
b = 2\mathbf{d}^T \boldsymbol{\Sigma}^{-1}(\mathbf{o}-\boldsymbol{\mu}), \\
c &= (\mathbf{o}-\boldsymbol{\mu})^T \boldsymbol{\Sigma}^{-1}(\mathbf{o}-\boldsymbol{\mu}) - \tau.
\end{aligned}
\]
The closest valid hit is given by the smaller positive root.
For our architectural study, this formulation exposes native ray--Gaussian intersection without polyhedral approximation.
Its arithmetic is dominated by symmetric matrix-vector products, dot products, and a quadratic solve, directly reflecting the workload that the hardware must execute.
Accordingly, the rest of this paper adopts this RaySplats-style workload model: rays are intersected directly with Gaussian primitives, and their contributions are accumulated in ray order.

\subsection{Why Hybrid Rendering Matters}
Once Gaussian rendering is expressed in a ray-tracing framework, it inherits many of the benefits of ray tracing.
Prior work shows that ray-traced 3DGS can naturally support reflections, refractions, shadows, stochastic ray sampling, highly distorted cameras, and rolling-shutter effects~\cite{moenneLoccoz3DGRT,wu3DGUT,byrskiRaySplats}.
It also becomes easier to combine Gaussian scene content with triangle-based geometry, mesh proxies, or standard graphics assets within the same rendering pipeline.
For instance, 3DGRT demonstrates compatibility with ray-based visual effects and mesh interactions, 3DGUT performs hybrid rendering with splatted primary rays and traced secondary rays, and RaySplats shows that Gaussian scenes can be rendered together with mesh-based objects and lighting effects.
These capabilities make mixed primitive support relevant even when Gaussian rendering remains the dominant workload: the architecture must handle the less frequent triangle path without duplicating an entire rendering backend.

Hybrid scenes therefore create a genuine mixed-intersection workload: different rays may encounter Gaussian primitives, triangles, or both.
Figure~\ref{fig:uie-mixed-rendering-stats} later quantifies this behavior for perspective, fisheye, and mesh-augmented rendering.
We treat this workload as an architectural requirement for unified ray--Gaussian and ray--triangle execution.

%% file: sections/03-Algorithm.tex
\begin{figure*}[th]
    \centering
    \begin{minipage}[t]{0.24\textwidth}
        \centering
        \includegraphics[width=\linewidth]{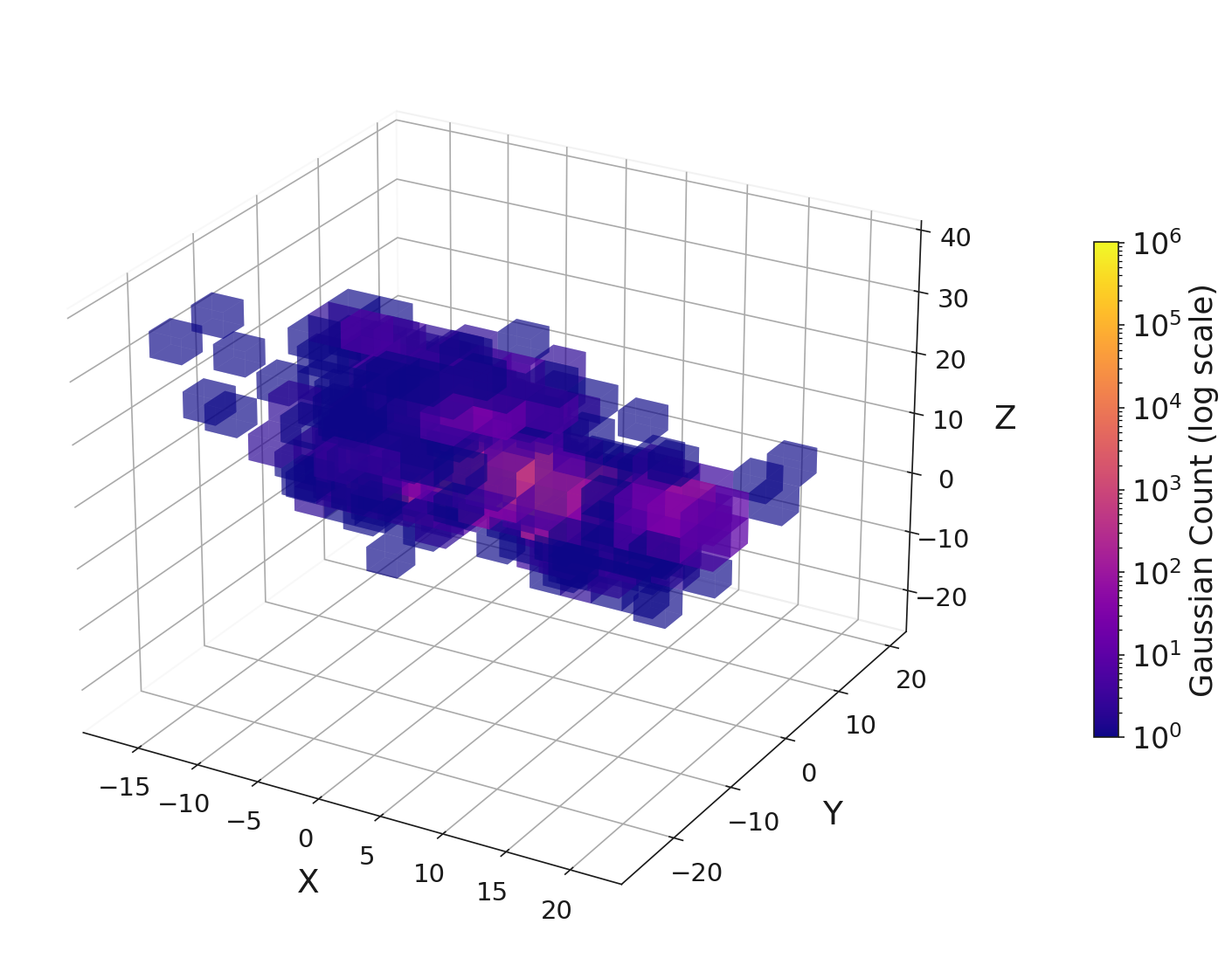}\\[-1mm]
        \footnotesize \textit{Kitchen}
    \end{minipage}\hfill
    \begin{minipage}[t]{0.24\textwidth}
        \centering
        \includegraphics[width=\linewidth]{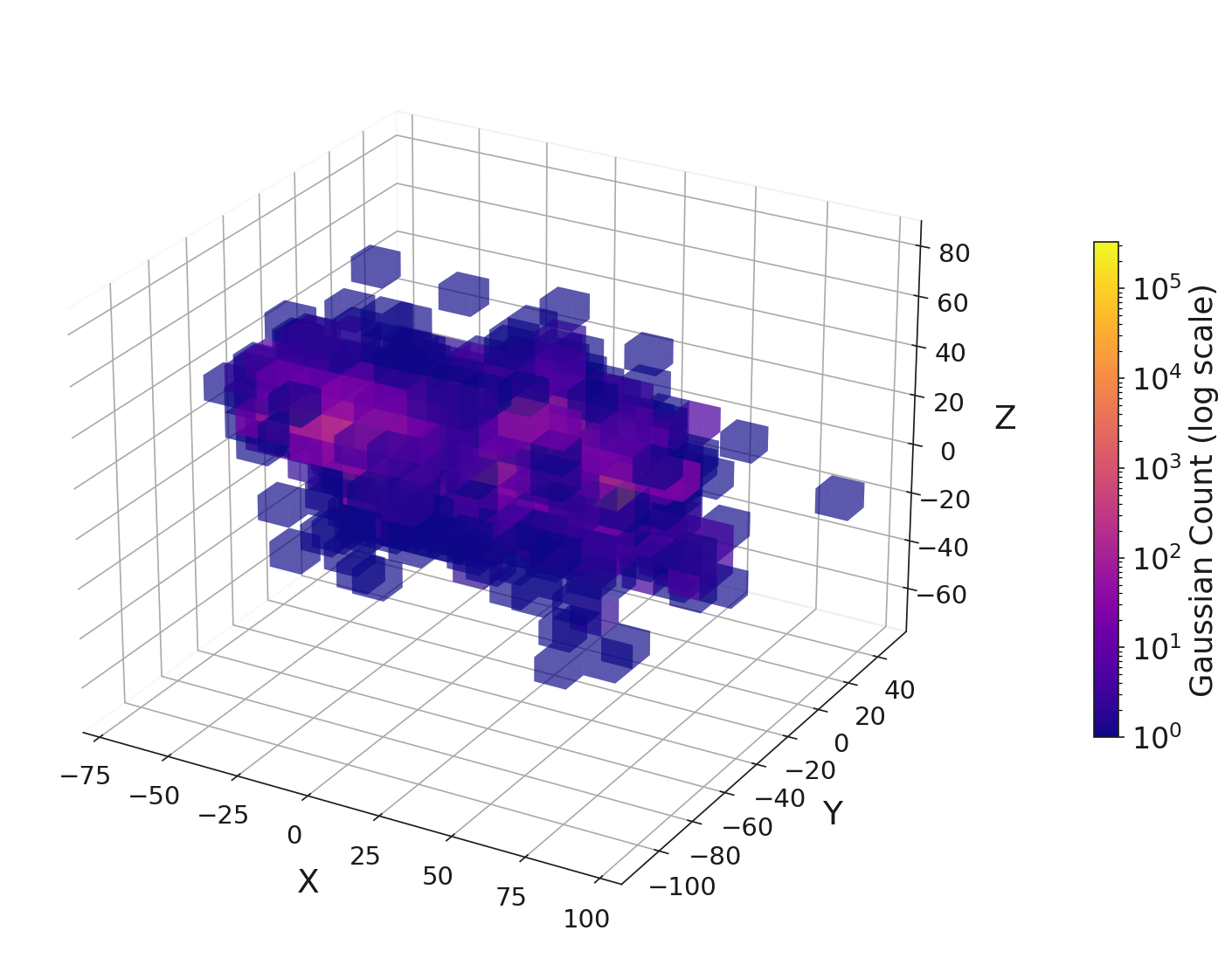}\\[-1mm]
        \footnotesize \textit{Stump}
    \end{minipage}\hfill
    \begin{minipage}[t]{0.24\textwidth}
        \centering
        \includegraphics[width=\linewidth]{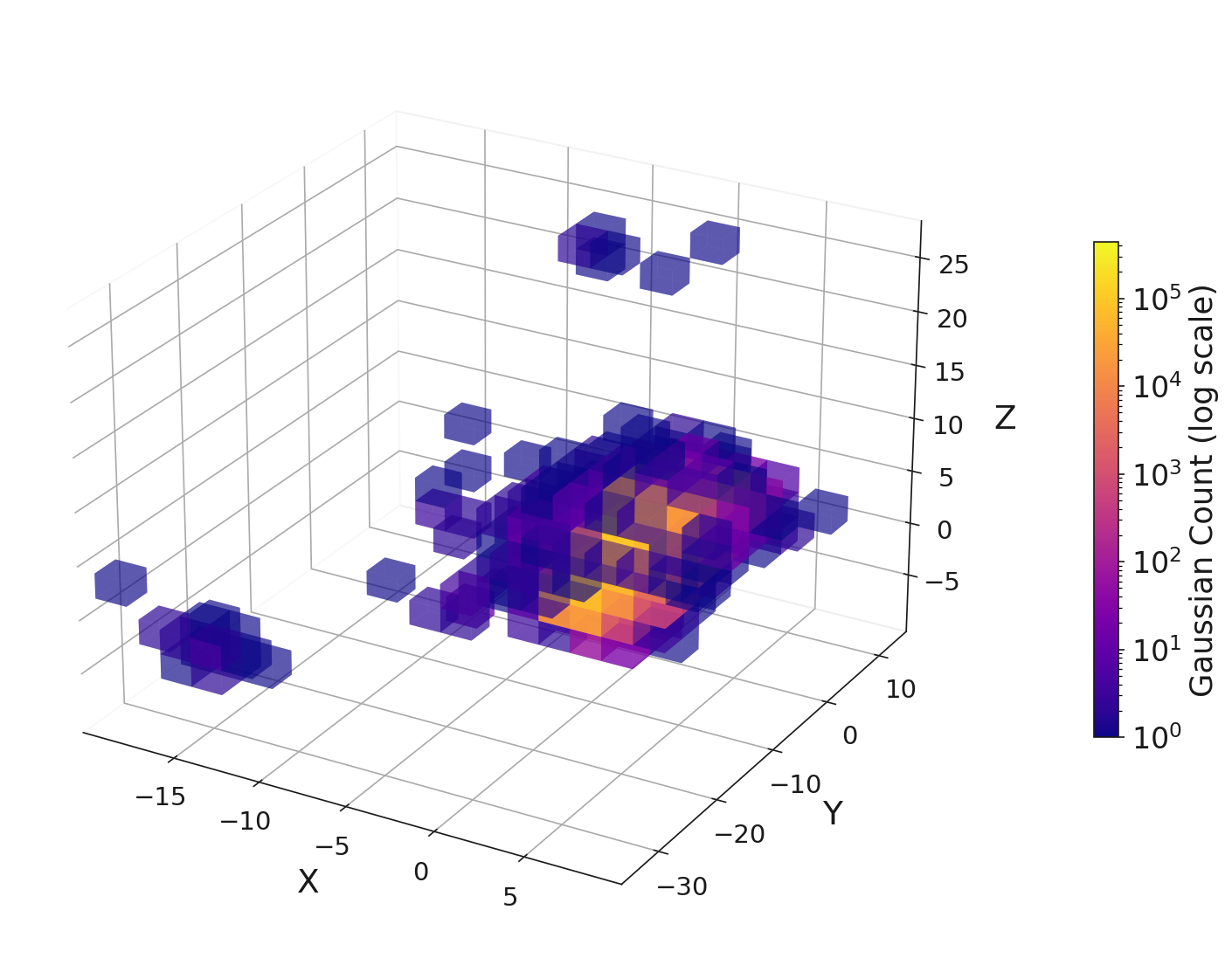}\\[-1mm]
        \footnotesize \textit{drJohnson}
    \end{minipage}\hfill
    \begin{minipage}[t]{0.24\textwidth}
        \centering
        \includegraphics[width=\linewidth]{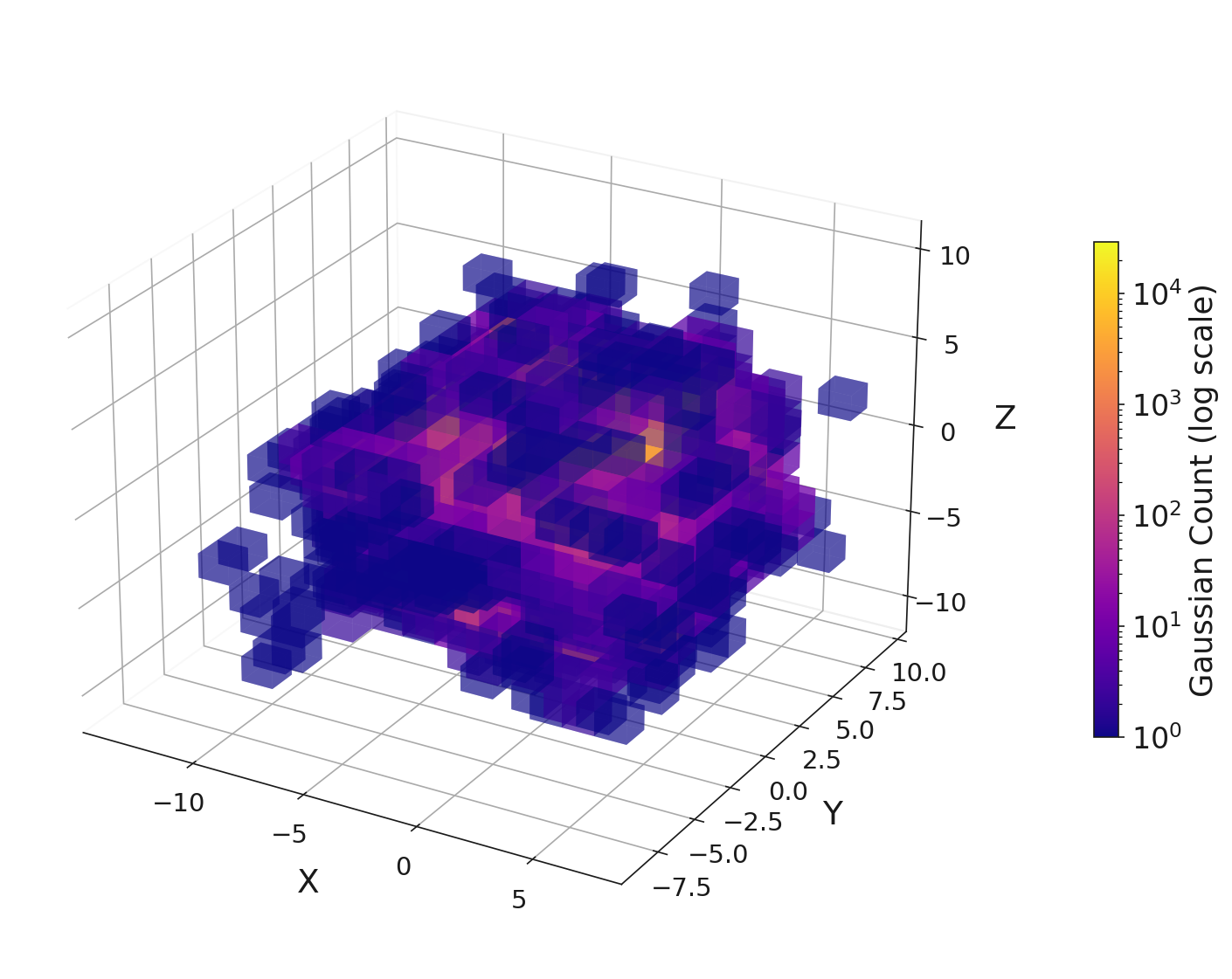}\\[-1mm]
        \footnotesize \textit{Playroom}
    \end{minipage}
    \caption{
        Gaussian-center density under a uniform $16^3$ grid.
        \textit{Kitchen} and \textit{Stump} are high-CV scenes; \textit{drJohnson} and \textit{Playroom} are lower-CV scenes.
    }
    \label{fig:motivation-imbalance}
\end{figure*}

\section{Motivation}

We characterize a RaySplats-style renderer at the scene, ray, and intersection levels.
Its workload is shaped by spatial imbalance, redundant neighboring-ray traversal, and scene-dependent Gaussian--triangle mixtures, motivating the three components of \HhRay.

\input{tables/motivation-grid-stats}

\subsection{Scene-Level Challenge: Spatial Imbalance in Gaussian Distributions}

Unlike mesh scenes, 3DGS Gaussians cluster around visible surfaces, leaving empty or weakly observed regions sparsely populated.
The skew varies by scene: the coefficient of variation in Table~\ref{tab:motivation-grid-stats} ranges from 2.7 (\textit{Playroom}) to 11.2 (\textit{Kitchen}).

This spatial skew is problematic for a conventional monolithic BVH, as shown in Figure~\ref{fig:motivation-imbalance}. A single global hierarchy must serve dense regions with long traversal and intersection queues as well as sparse regions that underutilize ray-tracing hardware.
This produces uneven per-ray cost, reducing parallel efficiency and complicating load balance as scenes grow.
The imbalance also makes a single global hierarchy a poor unit for memory placement and scheduling: dense regions dominate queues and cache traffic, while sparse regions leave traversal capacity idle.

Table~\ref{tab:motivation-grid-stats} quantifies this imbalance across all eight scenes under a uniform $8^3$ partition. On average, 69.8\% of grids are empty and another 20.9\% are sparse (1--99 Gaussians), leaving only 9.3\% dense (at least 100 Gaussians).
Even among non-empty grids, Gaussian counts remain highly uneven (average CV of 6.7).

\begin{takeawaybox}
\textbf{Takeaway.} Ray-traced 3DGS exhibits severe spatial imbalance. LB-SBVH restructures this workload into more uniform traversal units.
\end{takeawaybox}

\begin{figure}[t]
    \centering
    \includegraphics[width=\linewidth]{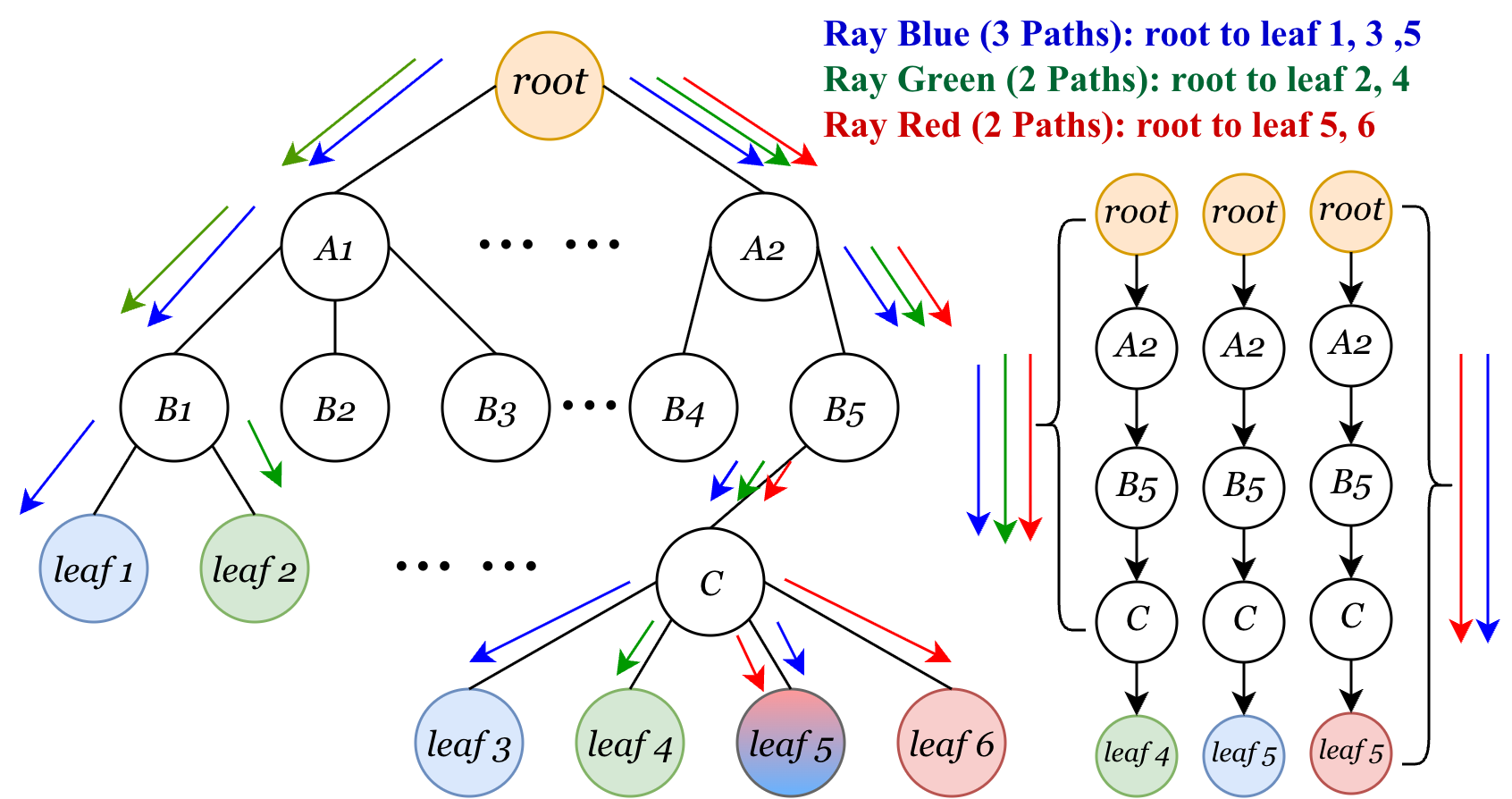}
    \caption{
        Illustration of path reuse in BVH traversal.
        Multiple neighboring rays share long traversal prefixes
        (e.g., \textit{root} $\rightarrow$ A2 $\rightarrow$ B5 $\rightarrow$ C) before diverging toward different leaf nodes, where each leaf node
        represents a Gaussian primitive.
        By reusing the shared prefix computed by a proxy ray,
        subsequent rays avoid redundant traversal of identical nodes.
    }
    \label{fig:motivation-traversal-sharing}
\end{figure}

\begin{figure*}[t]
    \centering
    \begin{minipage}[b]{0.36\textwidth}
        \centering
        \includegraphics[width=\textwidth]
        {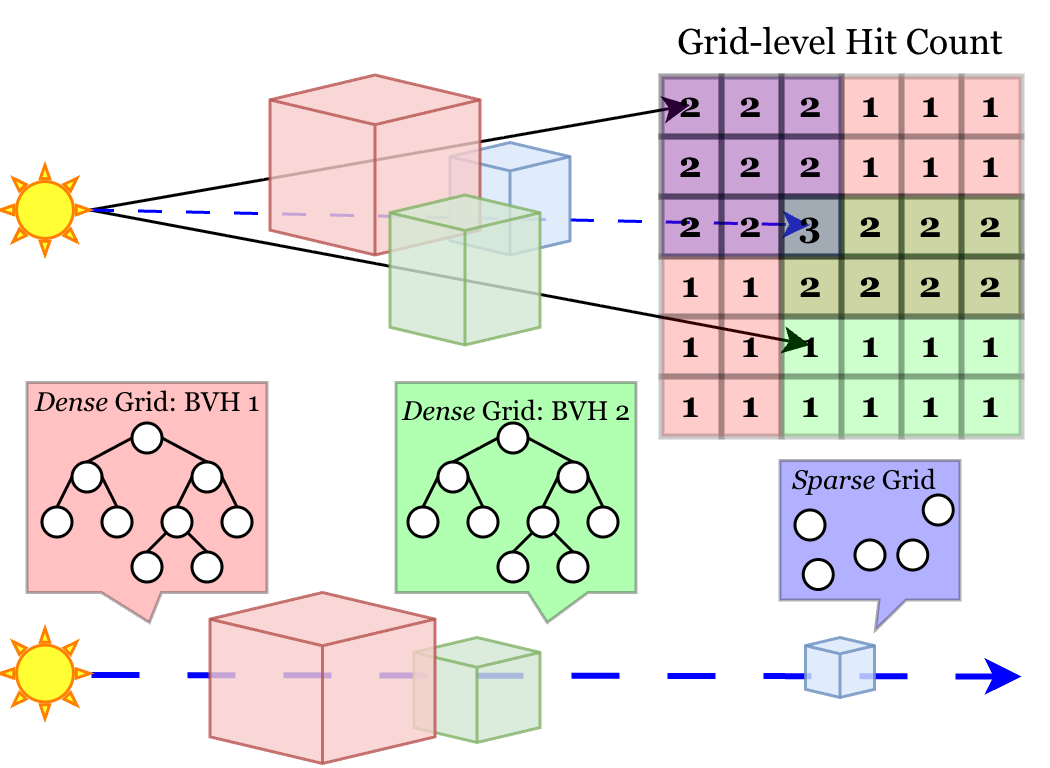}

        \vspace{2pt}
        \footnotesize
        (a) LB-SBVH density-aware partitioning
    \end{minipage}
    \hfill
    \begin{minipage}[b]{0.61\textwidth}
        \centering
        \includegraphics[width=\textwidth]
        {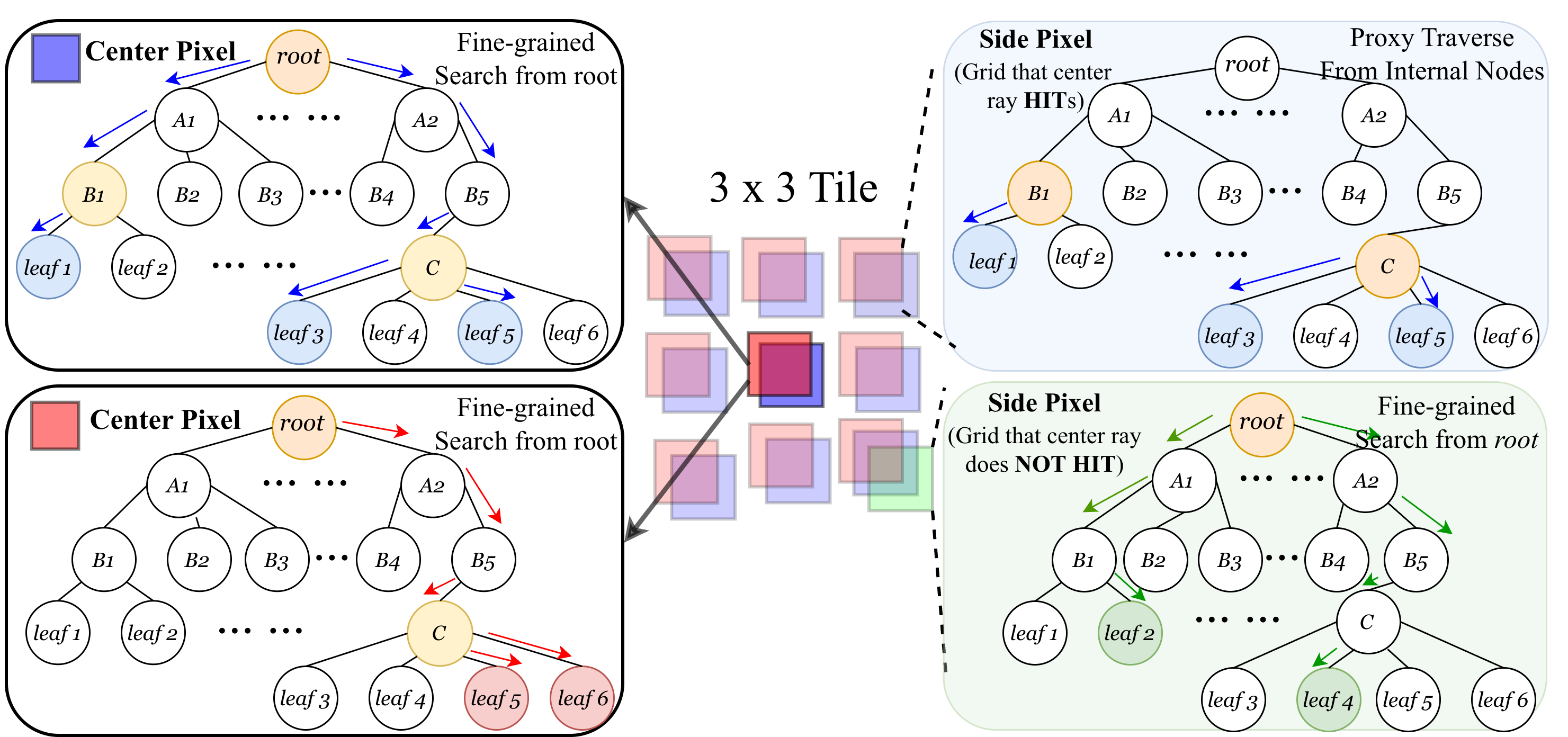}

        \vspace{2pt}
        \footnotesize
        (b) Ray-Cluster Path Reuse (RCPR) execution flow
    \end{minipage}

    \caption{
        Key algorithmic mechanisms of \HhRay.
        \textbf{(a)}~LB-SBVH partitions the scene into uniform
        spatial grids, each classified as \textit{Dense}
        (assigned a dedicated local BVH) or \textit{Sparse}
        (stored as a flat primitive list).
        \textbf{(b)}~RCPR designates the center ray of each
        $3 \times 3$ tile as a proxy that performs full BVH
        traversal; neighboring side rays short-circuit to the
        shared interior node when the grid matches, falling back
        to root traversal only for divergent grids.
    }
    \label{fig:lbsbvh-rcpr}
\end{figure*}

\subsection{Ray-Level Challenge: Redundant Traversal Across Neighboring Rays}
A second inefficiency remains even with a fixed acceleration structure: neighboring rays often share BVH prefixes because adjacent pixels enter the same coarse Gaussian regions before diverging near boundaries (Figure~\ref{fig:motivation-traversal-sharing}). This image-space coherence is especially strong at high resolution and in primary visibility.

Current execution nevertheless fetches and tests these upper-level nodes repeatedly, even though much of the traversal state is shared. This wastes memory bandwidth and RT Core cycles in coherent workloads such as mild lens distortion and some secondary-ray queries.
The wasted work is not limited to duplicate node tests; repeated address generation and state updates also increase pressure on the traversal stack and memory system.

\begin{takeawaybox}
\textbf{Takeaway.} Ray-traced 3DGS contains substantial inter-ray reuse. RCPR amortizes BVH traversal over a ray cluster instead of paying the full cost for every ray.
\end{takeawaybox}

\subsection{Intersection-Level Challenge: Heterogeneous Mixed-Primitive Intersection Workloads}
The final bottleneck appears when Gaussian rendering extends to hybrid scenes with triangle meshes, proxy geometry, or conventional ray-traced effects, where rays require Gaussian tests, triangle tests, or both.

A rigid pipeline specialized for one primitive type is therefore inefficient: a triangle-centric datapath loses efficiency when Gaussian intersections dominate, while a Gaussian-only datapath cannot serve triangle work. The challenge is both mixed execution and a dominant primitive that varies with scene, camera, and task.

\begin{takeawaybox}
\textbf{Takeaway.} Mixed workloads require an adaptive backend. UIE shares floating-point units between Gaussian and triangle paths and selects the active mode per test.
\end{takeawaybox}

%% file: tables/motivation-grid-stats.tex
\begin{table}[t]
\centering
\caption{Sub-region occupancy under an $8^3$ grid.}
\label{tab:motivation-grid-stats}

\small
\setlength{\tabcolsep}{4pt}

\begin{tabular}{lrrrr}
\toprule
\textbf{Scene} & \textbf{Empty} & \textbf{Sparse$^{*}$} & \textbf{Dense$^{*}$} & \textbf{CV} \\
\midrule
Bicycle   & 73.8\% & 18.2\% & 8.0\%  & 6.9  \\
Bonsai    & 56.6\% & 26.6\% & 16.8\% & 8.3  \\
Garden    & 75.8\% & 16.2\% & 8.0\%  & 6.2  \\
Kitchen   & 73.6\% & 21.7\% & 4.7\%  & 11.2 \\
Room      & 70.7\% & 21.9\% & 7.4\%  & 4.5  \\
Stump     & 65.6\% & 25.2\% & 9.2\%  & 10.1 \\
drJohnson & 88.3\% & 8.4\%  & 3.3\%  & 3.4  \\
Playroom  & 54.1\% & 29.3\% & 16.6\% & 2.7  \\
\midrule
\textbf{Avg.} & \textbf{69.8\%} & \textbf{20.9\%} & \textbf{9.3\%} & \textbf{6.7} \\
\bottomrule
\end{tabular}

\vspace{1pt}
\footnotesize
$^{*}$ Empty, Sparse, and Dense sum to 100\% of all grids. For a grid containing $N$ Gaussian centers, \emph{empty} denotes $N=0$, \emph{sparse} denotes $0<N<100$, and \emph{dense} denotes $N\geq100$. CV is computed over non-empty grids.

\end{table}

%% file: sections/04-Codesign.tex
\section{\texorpdfstring{\HhRay}{\HhRayPlain} Acceleration Framework}

\HhRay combines three mechanisms: LB-SBVH for scene partitioning, RCPR for traversal reuse, and UIE for mixed-primitive intersection. Together, they form a hybrid and hierarchical ray-tracing framework for efficient 3D Gaussian rendering.

\subsection{The End-to-End Rendering Pipeline}
The \HhRay pipeline reorganizes conventional ray tracing into four density-aware stages:

\paragraph{Raster-Assisted Culling}
Rather than starting each ray from a cold BVH traversal,
\HhRay first runs a lightweight low-resolution rasterization pass
to generate a \textit{grid hit map} that records per-grid hit counts (Figure~\ref{fig:lbsbvh-rcpr}).
The map prunes empty regions and skips rays with negligible Gaussian energy,
reducing initial traversal pressure.

\paragraph{Balanced Scene Partitioning}
The scene is partitioned into uniform spatial grids by \textit{LB-SBVH},
with each grid classified as dense or sparse by primitive count.
Dense grids are assigned dedicated local BVHs, while sparse grids store
primitives in flat lists for brute-force enumeration.
Culling then dispatches rays only to active grids, concentrating traversal
on occupied regions with bounded local workloads.

\paragraph{Coherent Path Traversal}
For each ray cluster, the \textit{RCPR} unit selects a
\textit{proxy ray} that performs BVH node fetching and internal
intersection tests. Its traversal state is broadcast to
worker rays, amortizing memory bandwidth across the group.

\paragraph{Adaptive Intersection \& Accumulation}
When rays reach leaf nodes, the \textit{UIE} selects the appropriate
Gaussian or triangle mode for each primitive, using a shared pool of floating-point units with mode-multiplexed inputs.
The pipeline then performs front-to-back alpha blending and color
accumulation.

\begin{figure*}[t]
    \centerline{\includegraphics[width=0.9\textwidth]{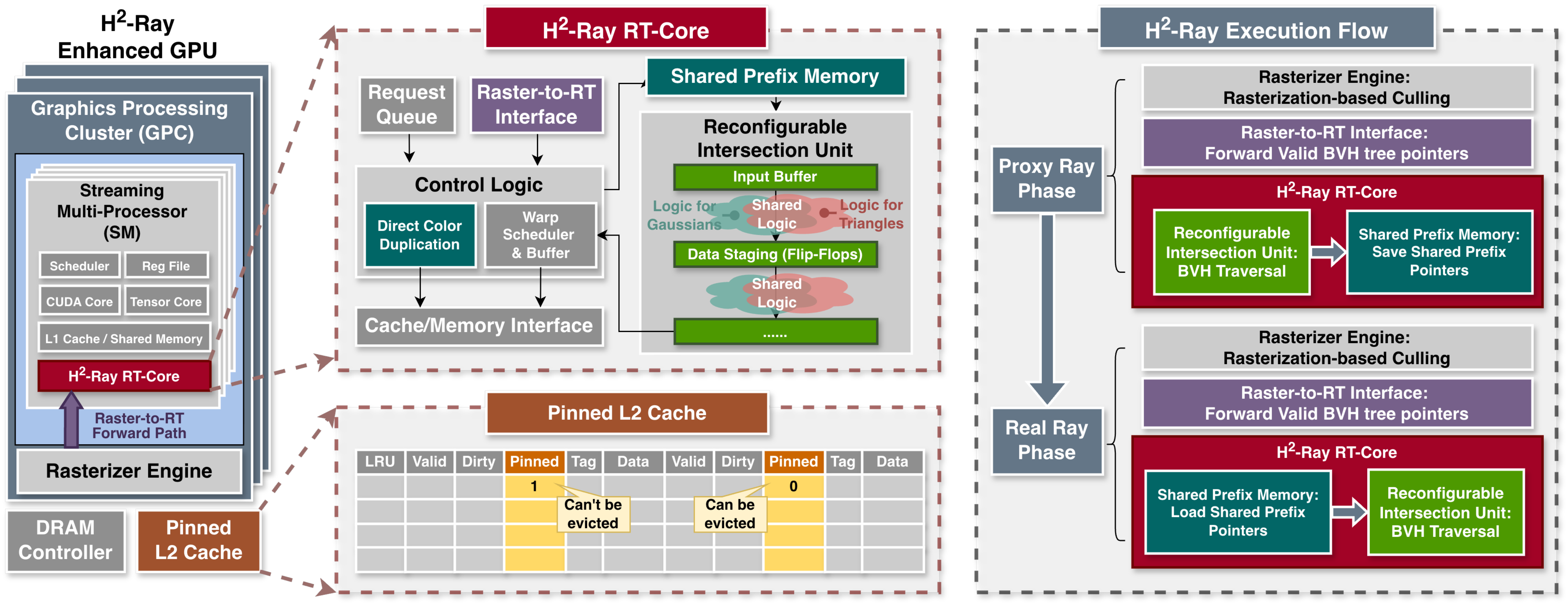}}
    \caption{
        Overall \HhRay hardware architecture.
        The system view shows GPC-level integration, while R2RFP, SPM, pinned cache, and UIE integrate partitioning, path reuse, and mixed-primitive intersection into the GPU pipeline.}
    \label{fig:overall_arch}
\end{figure*}

\subsection{Load-Balanced Spatial BVH (LB-SBVH)}
\label{subsec:lbsbvh}

3D Gaussian scenes are highly non-uniform, making a monolithic BVH prone to divergence and hotspots. The Load-Balanced Spatial BVH (LB-SBVH) converts this global workload into locally balanced traversal units (Figure~\ref{fig:lbsbvh-rcpr} (a)).

\paragraph{Density-Aware Sub-Region Construction}
LB-SBVH partitions the scene into a uniform spatial grid and chooses a representation per grid.
\textbf{Dense grids} ($N\geq100$ Gaussians) construct specialized shallow BVHs to accelerate traversal in regions with high primitive density.
\textbf{Sparse grids} ($0<N<100$ Gaussians) bypass BVH construction and store primitives in flat lists for direct enumeration, avoiding unnecessary tree overhead. Empty grids ($N=0$) are excluded from both categories.

\paragraph{Coverage-Preserving Grid Bounds}
Ownership and traversal bounds are separate.
Each Gaussian is assigned once by its center, avoiding duplicate occupancy counts.
Each non-empty grid then builds a tight conservative AABB around the full support of its assigned Gaussians.
These AABBs may expand beyond or shrink within the original cells and may overlap one another, so any ray intersecting a Gaussian support reaches its owning grid even when it misses the center cell.
The local BVH or sparse list performs the exact intersection, ensuring complete coverage without duplicating Gaussians.

\paragraph{Hardware Support for LB-SBVH}
LB-SBVH adds two hardware mechanisms. \textbf{Hybrid Metadata Storage (HMS)} combines a meta-tree with meta-bitmaps to encode BVH addresses and sub-region positions, allowing threads to locate target trees with little decoding overhead. The \textbf{Raster-to-RT Forward Path (R2RFP)} forwards BVH pointers directly from the Raster Engine to the RT Cores, avoiding DRAM synchronization and redundant top-level traversal.

\begin{takeawaybox}
\textbf{Summary.}
LB-SBVH addresses the scene-level challenge by partitioning an imbalanced Gaussian distribution into local traversal regions with bounded cost. HMS and R2RFP provide the metadata and dispatch support needed to realize this partitioning efficiently in hardware.
\end{takeawaybox}

\subsection{Ray-Cluster Path Reuse (RCPR)}

For primary and coherent secondary rays, neighboring paths through the BVH are highly correlated. RCPR shares intermediate states within localized $3 \times 3$ ray tiles (Figure~\ref{fig:lbsbvh-rcpr} (b)).

\paragraph{Proxy-Ray-Based Cluster Traversal}
For each tile, the center \textit{proxy ray} performs a full BVH traversal. When it reaches a leaf, hardware ``hops back'' by $c$ levels (e.g., $c=2$ or $3$) to define a \textit{coherence volume}. The eight worker rays are dispatched to these intermediate nodes, bypassing redundant upper-level tests.

\paragraph{Hardware Support for Path Reuse}
RT Cores are augmented with a \textbf{Shared Prefix Memory (SPM)} and a \textbf{Pinned Cache}. SPM stores the deepest nodes found by the proxy ray so worker rays resume from the longest shared path. Frequently accessed Gaussian primitives are pinned in L2 with a dedicated flag bit. A specialized internal Load-Store Unit (LSU) provides \textbf{Direct Color Duplication}, offloading copy-heavy work that would otherwise stall the warp.

\begin{takeawaybox}
\textbf{Summary.}
RCPR addresses the ray-level challenge by converting inter-ray coherence into reduced traversal and memory overhead. SPM, pinned caching, and direct color duplication provide the hardware support needed to exploit shared BVH prefixes efficiently.
\end{takeawaybox}

\subsection{Unified Intersection Engine (UIE)}

Since the Gaussian-to-triangle ratio varies by view, scene, and ray effect, fixed hardware for one primitive type is inefficient. UIE reuses the LB-SBVH grid structure: each grid points to either a Gaussian-BVH or a triangle-BVH, keeping primitive selection local to leaf-level intersection while ray generation, culling, and blending remain unchanged.

\paragraph{Structural Isomorphism Between Intersection Pipelines}
Gaussian ellipsoid intersection and M\"oller-Trumbore triangle intersection follow a similar algebraic flow: both transform the ray into a local frame, execute multiply-accumulate stages, and solve for the intersection distance~$t$. Many floating-point units can therefore be \textit{shared}; switching primitive types changes operands and control bits rather than the pipeline topology.

\paragraph{Reconfigurable Unified Datapath}
We instantiate \UIETotalFPUnits{} floating-point units organized in an eight-stage feed-forward pipeline.
Of these, \UIESharedFPUnits{} units (\UIESharedFPPct\%) are \textit{shared} between Gaussian and triangle modes: they execute the same operation in both modes but receive different operands via mode-MUX input selection.
The remaining units are dedicated to one primitive type: \UIEGaussianFPUnits{} units (\UIEGaussianFPPct\%) serve only the Gaussian path, while \UIETriangleFPUnits{} units (\UIETriangleFPPct\%) serve only the triangle path.
The Gaussian path needs more dedicated hardware because its quadratic solve and alpha computation have no triangle analog, while M\"oller--Trumbore is more compact.
Table~\ref{tab:uie-fp-sharing} reports sharing by unit type: multipliers and reciprocal units are shared most aggressively, while adders and the square-root unit are largely path-specific.

\input{tables/uie-fp-sharing}

The active datapath is selected per intersection by the primitive-type field in the ray-intersection instruction; no runtime workload prediction is required. Once traversal reaches a leaf, UIE gates the corresponding operands into the shared pipeline, so primitive-type selection requires no software branching while the triangle path stays lightweight. Triangle-dedicated units account for \UIETriangleFPUnits{} of the \UIETotalFPUnits{} floating-point units (approximately \UIETriangleFPPct\%); this is a unit-count fraction, not an area-overhead estimate.

\begin{takeawaybox}
\textbf{Summary.}
UIE addresses the intersection-level challenge with a reconfigurable datapath for Gaussian and triangle primitives. This shared structure avoids dedicating separate hardware to each primitive type in mixed-primitive workloads.
\end{takeawaybox}

\begin{figure}[t]
    \centering
    \begin{minipage}{0.55\columnwidth}
        \centering
        \includegraphics[width=\linewidth]{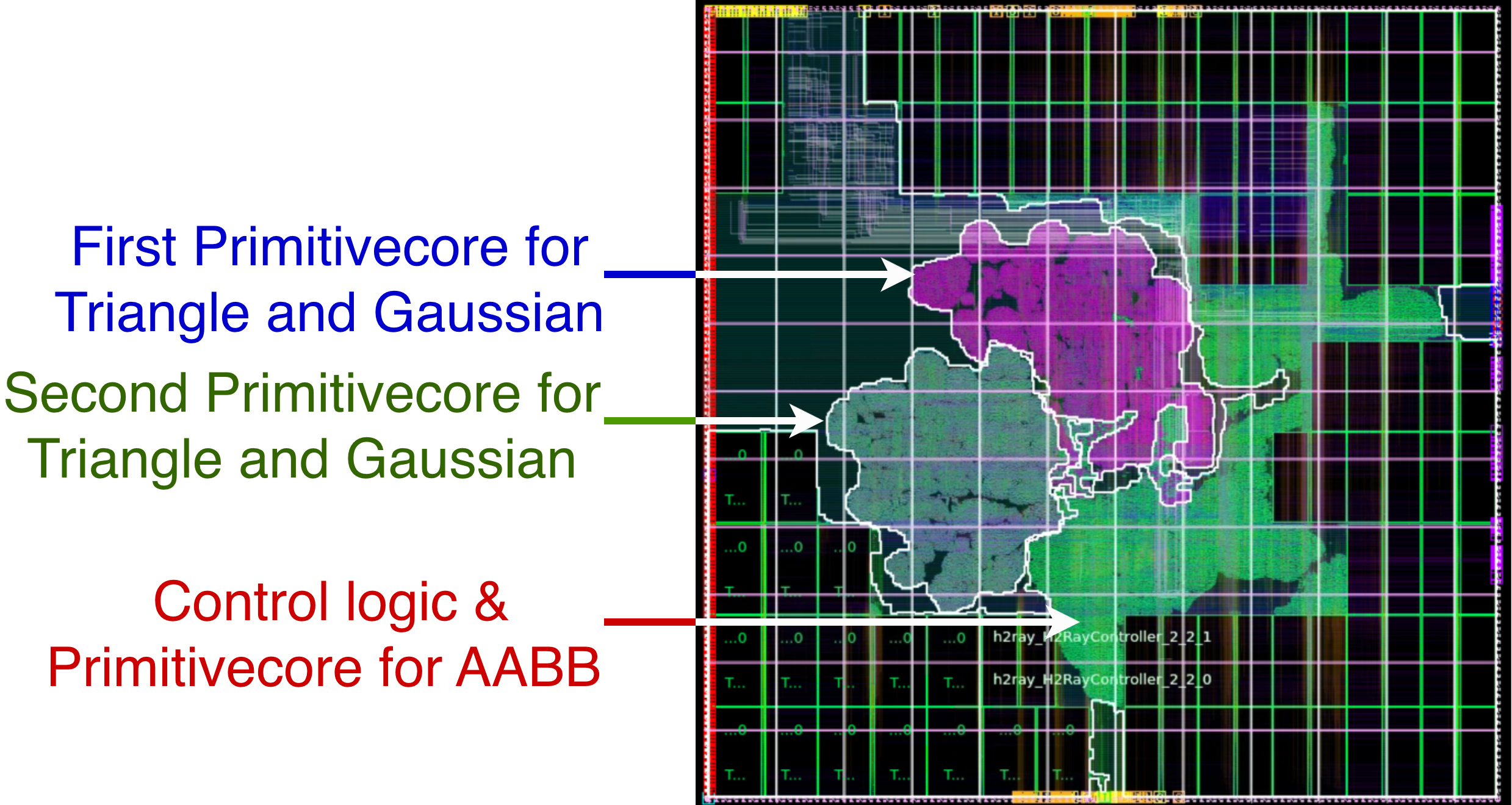}
    \end{minipage}
    \hfill
    \begin{minipage}{0.42\columnwidth}

    \footnotesize
    \setlength{\tabcolsep}{3pt}

    \resizebox{\linewidth}{!}{%
    \begin{tabular}{lc}

    \toprule

    \textbf{Metric / Core}
    & \textbf{Value} \\

    \midrule

    Technology Node & 28 nm \\
    Core Area & $0.81\ \text{mm}^2$ \\
    Power & $24.7\ \text{mW}$ \\
    PrimitiveCores & $2$ \\
    AABB Units & $2$ \\
    Peak FLOPs/cycle & $213$ \\

    \bottomrule

    \end{tabular}%
    }
    \end{minipage}
    \caption{
    Layout of the proposed \HhRay core after synthesis, placement, and routing in a 28\,nm process.
    The table summarizes the hardware configuration of a single core.
    Power is scaled to the 1\,GHz operating point from post-synthesis measurements at 100\,MHz, assuming dynamic power scales linearly with clock frequency while leakage power remains constant.
    }

    \label{fig:chipdesign}
\end{figure}

\subsection{Hardware Integration}

Figure~\ref{fig:overall_arch} shows how LB-SBVH, RCPR, and UIE are integrated into the GPU pipeline. At the system level (\textit{left}), each Graphics Processing Cluster (GPC) contains a Streaming Multiprocessor (SM), an \HhRay RT Core, and a Raster Engine connected by R2RFP. The Pinned L2 Cache sits beside the DRAM controller to support RCPR. This organization leaves shader execution, memory hierarchy, and rasterization largely intact; \HhRay adds only the paths needed to pass culling and reuse information to the RT Core. This section describes the target hardware path; the evaluation separately reports modeled RT Core behavior and measured CUDA kernels, as detailed in Section~\ref{sec:eval}.

The \HhRay RT Core (\textit{center}) adds four blocks to a conventional RT Core: (1)~a \textit{Raster-to-RT Interface} that receives BVH pointers without a DRAM round-trip, (2)~a \textit{Shared Prefix Memory} (SPM) for proxy-ray path pointers, (3)~a \textit{Reconfigurable Intersection Unit} that switches between Gaussian and triangle logic, and (4)~a \textit{Direct Color Duplication} unit that offloads pixel-copy operations from the SM. These additions are placed around the existing traversal and intersection flow: R2RFP supplies the starting BVH handle, SPM changes where worker rays resume traversal, UIE changes the leaf test, and color duplication handles repeated output writes.

Execution proceeds in two phases (\textit{right}). In the \textit{Proxy Ray Phase}, the Raster Engine performs low-resolution culling, forwards valid grid BVH handles through R2RFP, and the RT Core writes proxy-ray prefix pointers into SPM. In the \textit{Real Ray Phase}, worker rays load these pointers, skip redundant upper-level nodes, and proceed to the Reconfigurable Intersection Unit. The two phases therefore reuse traversal state while leaving the front-to-back blending semantics unchanged; the quality impact of this approximate reuse is evaluated in Section~\ref{sec:eval}. The Pinned L2 Cache (\textit{bottom}) protects marked cache lines from eviction so Gaussian primitives found during the proxy phase remain resident for worker rays.

%% file: tables/uie-fp-sharing.tex
\begin{table}[t]
\centering
\caption{FP unit classification in the UIE reconfigurable datapath.}
\label{tab:uie-fp-sharing}
\small
\begin{tabular}{@{}lrrrr@{}}
\toprule
\textbf{Unit Type} & \textbf{Total} & \textbf{Shared} & \textbf{Gaussian} & \textbf{Triangle} \\
\midrule
fp\_mult  & 30 & 18 (60\%) & 9  & 3 \\
fp\_add   & 14 & 2  (14\%) & 10 & 2 \\
fp\_sub   & 18 & 6  (33\%) & 3  & 9 \\
fp\_mac   & 13 & 4  (31\%) & 9  & 0 \\
fp\_recip & 2  & 1  (50\%) & 0  & 1 \\
fp\_sqrt  & 1  & 0         & 1  & 0 \\
\midrule
\textbf{Total} & \textbf{\UIETotalFPUnits} & \textbf{\UIESharedFPUnits (\UIESharedFPPct\%)} & \textbf{\UIEGaussianFPUnits (\UIEGaussianFPPct\%)} & \textbf{\UIETriangleFPUnits (\UIETriangleFPPct\%)} \\
\bottomrule
\end{tabular}

\vspace{2pt}
\footnotesize
Unit Type: \emph{fp\_mult} = multiplier, \emph{fp\_add} = adder, \emph{fp\_sub} = subtractor, \emph{fp\_mac} = multiply-accumulate unit, \emph{fp\_recip} = reciprocal unit, \emph{fp\_sqrt} = square-root unit.
\end{table}

%% file: sections/06-Evaluation.tex
\section{Evaluation}
\label{sec:eval}

\input{tables/dataset}

\input{tables/quality_stats_8scenes}

\input{tables/runtime_microbench_combo}

\subsection{Evaluation Settings}

\paragraph{Baselines.}
We use the official \textit{RaySplats}~\cite{byrskiRaySplats} implementation on an NVIDIA RTX A5000 as the baseline, using its standard preprocessing and training pipeline to generate all Gaussian PLY inputs. Because its Windows OptiX pipeline cannot directly support \HhRay, we port it to Linux and build a C++ validation renderer.
The same C++ H2-Ray renderer used to produce the \HhRay outputs cross-checks image formation against the original RaySplats implementation; the resulting matched-view comparison is reported in Table~\ref{tab:quality_stats}.

\paragraph{Datasets}
Table~\ref{tab:scene-summary} covers six Mip-NeRF 360 scenes~\cite{mipnerf360} and two Deep Blending scenes~\cite{deepblending}. Following the standard protocol, outdoor and indoor Mip-NeRF 360 scenes use $4\times$ and $2\times$ downsampling, respectively; for Deep Blending, we report the actual rendered resolution used rather than assuming a dataset-wide downsampling factor. Unless otherwise stated, view counts and multi-view quality statistics use the camera-name intersection for which ground truth, RaySplats, and \HhRay outputs are all available. These evaluation views are matched rendered cameras, not a held-out test split, and can overlap with views used to train the RaySplats scene representation. Runtime and visual-preview results use fixed representative views for per-scene illustration; they are not assumed to be the same camera as one another or as the views used for multi-view statistics.

\paragraph{Evaluation Platform Setup}
We combine cycle-accurate simulation for the modified RT Core with GPU measurements for unmodified CUDA execution. Simulation is limited to the proposed RT Core traversal and intersection datapaths that do not exist in the baseline GPU. All CUDA-core work, including intersection, sorting, and auxiliary kernels, is measured on the RTX A5000. The synthesizable C++/SystemC core is translated to RTL with Siemens Catapult HLS~\cite{siemens_catapult_hls}, then synthesized, placed, and routed at 28\,nm, 0.9\,V, and 1\,GHz with Synopsys Fusion Compiler~\cite{synopsys_fusion_compiler}; Figure~\ref{fig:chipdesign} shows its layout. Power is extrapolated to 1\,GHz from post-synthesis results at 100\,MHz.

We scale to the A5000's \AFiveKRTTFLOPS{} RT TFLOPS: 255 cores at 213 RT-equivalent FP32 operations/cycle and 1\,GHz provide \HhRayScaledRTTFLOPS{} RT TFLOPS.

\paragraph{Metric Definitions.}
Runtime results in Figure~\ref{fig:runtime-time-results} report per-frame latency for representative views. Each scene entry is a profiled run after initialization, without additional warm-up or repeated-run averaging. The arithmetic/geometric averaging conventions for the runtime table are given in its caption.
The timing excludes offline RaySplats training/preprocessing, model loading, and image I/O; it covers the rendering pipeline components shown in Figure~\ref{fig:runtime-time-results}(a).
Quality results use the camera-name intersection in Table~\ref{tab:scene-summary}, so every reported view has ground truth, RaySplats, and \HhRay outputs.
PSNR is computed over RGB 8-bit images as $20\log_{10}(255/\mathrm{RMSE})$ using all pixels in the common top-left image region when two images differ by at most a boundary pixel.
SSIM is computed on RGB images, not grayscale, using a Gaussian window with $\sigma=1.5$, an $11\times11$ support, data range 255, and channel averaging.
Both PSNR and SSIM compare each rendered image against its ground-truth image; they are not computed between RaySplats and \HhRay outputs.

\paragraph{Performance and Workload Characterization}
We separate end-to-end runtime, the RT/CUDA latency breakdown, the RT-Core-only LB-SBVH microbenchmark, and primitive-test counts for perspective, fisheye, and hybrid rendering. The first two combine the modeled modified RT Core with measured CUDA kernels; the microbenchmark isolates simulated RT-Core traversal, while RCPR is examined separately through hop-distance quality sensitivity. The 3.77$\times$ end-to-end result is therefore a combined-system result, not an isolated attribution to LB-SBVH or RCPR. Per-scene runtime averages are arithmetic means, whereas the reported average speedup is the geometric mean of per-scene speedups.

\subsection{End-to-End Results}

\paragraph{Efficiency Analysis}
Figure~\ref{fig:runtime-time-results}(a) combines the modeled modified RT Core with measured CUDA execution and reports a $\EvalGeoSpeedup\times$ geometric-mean speedup over RaySplats. The arithmetic-mean runtime decreases from \RaySplatsAvgRuntimeMs{} to \HhRayAvgRuntimeMs\,ms, with per-scene speedups ranging from $\EvalMinSpeedup\times$ on \textit{Stump} to $\EvalMaxSpeedup\times$ on \textit{Garden}; \textit{drJohnson} and \textit{Playroom} reach $\DrJohnsonSpeedup\times$ and $\PlayroomSpeedup\times$.
The average runtime and average speedup use different means, so $\EvalGeoSpeedup\times$ is not the ratio of the two average runtime entries.

The same figure decomposes \HhRay latency. Since only each tile's center ray is traced on the RT Core, parallel CUDA work largely hides RT latency. The RT portion is estimated with the modified RT Core model, while CUDA intersection and sorting are measured on the RTX A5000. Intersection dominates CUDA latency in six scenes; sorting dominates in \textit{Stump} and \textit{Playroom}, where per-ray hit counts vary more. In this evaluated pipeline, worker-ray intersection and sorting run as CUDA kernels, so the end-to-end runtimes do not include any benefit from executing worker rays on the proposed Reconfigurable Intersection Unit.

\begin{figure*}[t]
    \centering
    \includegraphics[width=0.85\textwidth]{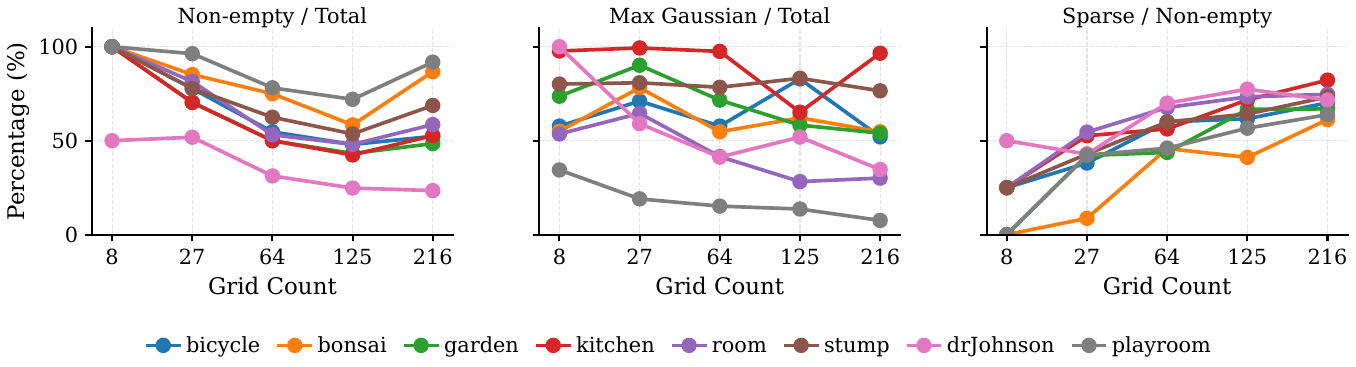}
    \caption{
        Impact of grid granularity on spatial workload distribution.
        Left: fraction of non-empty grids over total grids.
        Middle: fraction of Gaussians contained in the densest grid.
        Right: fraction of sparse grids ($0 < N < 100$) among non-empty grids.
    }
    \label{fig:grid-distribution}
\end{figure*}

\begin{figure}[t]
    \centering
    \includegraphics[width=\columnwidth]{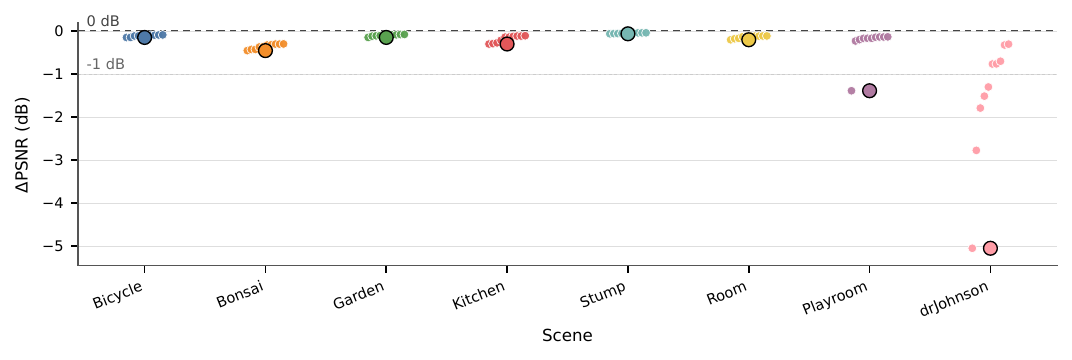}
    \caption{
        Worst-case per-view PSNR differences across \NumEvalScenes{} scenes.
        Each point denotes one of the ten lowest per-view differences between \HhRay and RaySplats for a scene.
        The black-outlined marker denotes the worst view in each scene.
        Values closer to zero indicate better agreement with RaySplats.
    }
    \label{fig:bottom10-delta-psnr}
\end{figure}

\paragraph{Quality Validation}
Table~\ref{tab:quality_stats} reports per-view PSNR and SSIM statistics on matched RaySplats and \HhRay views. The arithmetic mean of the per-scene average PSNR deltas is \QualityAvgPSNRDelta\,dB for \HhRay{} minus RaySplats, while the min/max statistics expose view-level outliers. Figure~\ref{fig:bottom10-delta-psnr} visualizes the ten worst per-view PSNR differences for each scene, and Figure~\ref{fig:scene-preview-8} provides representative examples. Unless noted, experiments use 2 hops and a $3^3$ grid.

For the two largest \textit{drJohnson} outliers listed in Table~\ref{tab:drjohnson-outliers}, we rerun the corresponding views with proxy reuse disabled, so every ray performs independent traversal, and also test hop distance 3.
The no-proxy reruns improve PSNR by less than 0.4\,dB for the inspected views, and the default proxy result is sometimes better. Since neither setting consistently closes the RaySplats--\HhRay gaps, the outliers are unlikely to be caused primarily by proxy reuse or hop truncation.
Because RaySplats does not expose the exact OptiX BVH construction and traversal details, the remaining gaps may instead stem from implementation-level differences such as BVH structure, traversal order, or floating-point ordering.

\input{tables/drjohnson-outliers}

\input{tables/sensitivity}
\input{tables/fisheye_object}

\subsection{Sensitivity Analysis}

\paragraph{Hop Distance Analysis.}
Table~\ref{tab:hop-distribution} reports PSNR for representative views across \NumEvalScenes{} scenes.
Unlike Table~\ref{tab:quality_stats}, which compares \HhRay against RaySplats over matched multi-view sets, this table compares RCPR hop settings against the C++ full per-ray traversal baseline on selected representative views.
Relative to that baseline, 0-, 1-, and 2-hop reuse reduce the across-scene mean by 2.99, 1.32, and \RCPRTwoHopAvgLoss\,dB, respectively; the maximum per-scene loss at 2 hops is \RCPRTwoHopMaxLoss\,dB.
These representative-view statistics do not bound all camera views, but they show the quality trend as the reuse window is relaxed.
At 0 hops, neighboring rays reuse only the center ray's leaf-level hits, which is vulnerable when a tile crosses a foreground/background boundary. This effect is most visible in \textit{Garden} and \textit{Playroom}, where 0-hop reuse loses 5.55 and 10.19\,dB, respectively.
One hop recovers much of the loss by resuming traversal from a higher shared state, but \textit{Playroom} still loses 6.94\,dB because its transitions remain difficult.
At 2 hops, all scenes are within \RCPRTwoHopMaxLoss\,dB of baseline on these views, so we use 2 hops as the accuracy--cost balance.

\paragraph{Grid Division Analysis.}

Figure~\ref{fig:grid-distribution} evaluates grid granularity, which changes traversal organization but not correctness because all candidate Gaussians remain reachable.
Grid statistics use the center-based ownership rule, while traversal uses the support-aware per-grid AABBs described in Section~\ref{subsec:lbsbvh}; Gaussians are not duplicated across cells.
Moving from coarse to moderate partitions generally reduces occupied-grid fraction and densest-cell concentration by separating empty regions from dense Gaussian clusters.
At finer settings, occupied- and sparse-grid fractions increase as dense regions are subdivided into cells with too few primitives to benefit from BVH pruning.
These trends motivate our $3^3$ default, which isolates major hotspots while retaining enough Gaussians per occupied grid for local BVHs.
Figure~\ref{fig:motivation-imbalance} uses $16^3$ to visualize spatial skew, Table~\ref{tab:motivation-grid-stats} uses $8^3$ for occupancy statistics, and the renderer uses $3^3$ by default.
Table~\ref{tab:motivation-grid-stats} reports sparse and dense fractions over all grids, whereas Figure~\ref{fig:grid-distribution} reports sparse grids over non-empty grids, so the percentages are not directly comparable.

Table~\ref{tab:runtime-microbench} isolates modified RT Core traversal by comparing one global BVH with 27 local grid BVHs.
Each scene row averages 40 benchmark patches over proxy center rays sent to the RT Core traversal engine; CUDA-only reused rays are excluded. All rows use the same per-ray correctness-checking procedure. The across-scene arithmetic means are \LBSBVHOneGridAvg{} and \LBSBVHTwentySevenGridAvg{} cycles per ray, while the geometric mean of the per-scene speedups is \LBSBVHAvgSpeedup$\times$.
Across \NumEvalScenes{} scenes, 27 grids reduce traversal latency by $\LBSBVHMinSpeedup\times$--$\LBSBVHMaxSpeedup\times$.
The benefit is largest for \textit{Bonsai} ($1.31\times$; 7,207 to 5,481 cycles per ray) and \textit{Kitchen} ($1.27\times$), and smallest for \textit{Room} ($1.08\times$), while remaining positive in every scene.
The 27-grid organization also reduces DRAM reads by \LBSBVHDRAMMin\%--\LBSBVHDRAMMax\%, with an arithmetic mean reduction of \LBSBVHDRAMAvg\% across scenes.
The \textit{Room} row was additionally checked for traversal correctness on representative views.
As the table reports only the modeled RT Core microarchitecture, it represents the traversal-level contribution of LB-SBVH rather than end-to-end runtime.

\subsection{Ray Tracing with Heterogeneous Primitives}
Figure~\ref{fig:uie-mixed-rendering-stats} characterizes perspective, fisheye, and hybrid mesh workloads supported by the same pipeline.
The four examples use the \textit{Room} scene at its evaluation resolution with the same acceleration structure and scheduling flow, while varying the camera model and mesh insertion.
They use the default 2-hop RCPR setting and $3^3$ grid partition.
Across all modes, Gaussian tests dominate the intersection workload.
Perspective-plus-mesh performs 1.90 billion ray--Gaussian tests but only 1.38 million ray--triangle tests, while fisheye-plus-mesh performs 1.75 billion and 0.73 million, respectively.
Triangle tests therefore remain a small fraction of the intersection workload.

The fisheye camera changes ray directions and coverage, reducing ray--AABB and ray--Gaussian counts relative to perspective rendering.
Adding meshes introduces triangle tests while leaving the Gaussian workload nearly unchanged (1.87 to 1.90 billion tests under perspective rendering).
These results justify the UIE design: the datapath supports both intersection types and shares arithmetic units with the less frequent triangle path. UIE consequently broadens the rendering modes supported by \HhRay without a separate triangle-only pipeline.
This experiment establishes the mixed-primitive workload need rather than an isolated UIE speedup or area reduction; the architecture section reports UIE's FP-unit sharing as a structural resource breakdown.

%% file: tables/dataset.tex
\begin{table}[t]
\caption{
Summary of the evaluation scenes. Resolution denotes the rendered image size used in our experiments.
Eval. Views denotes matched rendered cameras, not a held-out test split.
}
\label{tab:scene-summary}

\centering
\scriptsize
\setlength{\tabcolsep}{3pt}

\begin{tabular}{@{}llcccc@{}}

\toprule

\textbf{Scene}
& \textbf{Dataset}
& \textbf{Resolution}
& \textbf{Category}
& \textbf{\#Gaussians}
& \textbf{Eval. Views} \\

\midrule

\textit{Bicycle}
& Mip-NeRF 360
& $1236\times822$
& Outdoor
& 1.77M
& 169 \\

\textit{Bonsai}
& Mip-NeRF 360
& $1559\times1039$
& Indoor
& 1.22M
& 255 \\

\textit{Garden}
& Mip-NeRF 360
& $1297\times840$
& Outdoor
& 5.37M
& 161 \\

\textit{Kitchen}
& Mip-NeRF 360
& $1558\times1039$
& Indoor
& 1.69M
& 244 \\

\textit{Room}
& Mip-NeRF 360
& $1557\times1038$
& Indoor
& 1.40M
& 272 \\

\textit{Stump}
& Mip-NeRF 360
& $1244\times825$
& Outdoor
& 1.04M
& 109 \\

\midrule

\textit{drJohnson}
& Deep Blending
& $666\times438$
& Indoor
& 3.08M
& 230 \\

\textit{Playroom}
& Deep Blending
& $632\times416$
& Indoor
& 1.01M
& 196 \\

\bottomrule

\end{tabular}

\end{table}

%% file: tables/quality_stats_8scenes.tex
\begin{table*}[t]
\centering
\caption{
Per-view PSNR and SSIM comparison between RaySplats and \HhRay across \NumEvalScenes{} scenes.
$\Delta$ denotes \HhRay minus RaySplats.
}
\label{tab:quality_stats}
\resizebox{\textwidth}{!}{
\begin{tabular}{l r|rrr|rrr|rrr|rrr|rrr|rrr}
\toprule
\multirow{2}{*}{\textbf{Scene}} & \multirow{2}{*}{\textbf{Views}}
& \multicolumn{3}{c|}{\textbf{RaySplats PSNR}}
& \multicolumn{3}{c|}{\textbf{\HhRay PSNR}}
& \multicolumn{3}{c|}{\textbf{$\Delta$ PSNR}}
& \multicolumn{3}{c|}{\textbf{RaySplats SSIM}}
& \multicolumn{3}{c|}{\textbf{\HhRay SSIM}}
& \multicolumn{3}{c}{\textbf{$\Delta$ SSIM}} \\
& & Min & Max & Avg & Min & Max & Avg & Min & Max & Avg & Min & Max & Avg & Min & Max & Avg & Min & Max & Avg \\
\midrule
\textit{Bicycle} & 169 & 16.16 & 31.07 & 24.24 & 16.16 & 31.01 & 24.22 & -0.14 & 0.04 & -0.02 & 0.3564 & 0.9145 & 0.7633 & 0.3561 & 0.9143 & 0.7626 & -0.0018 & 0.0001 & -0.0008 \\
\textit{Bonsai} & 255 & 24.68 & 35.77 & 32.87 & 24.62 & 35.75 & 32.85 & -0.45 & 0.48 & -0.02 & 0.9018 & 0.9664 & 0.9487 & 0.9005 & 0.9664 & 0.9474 & -0.0039 & 0.0008 & -0.0013 \\
\textit{Garden} & 161 & 24.73 & 31.62 & 29.39 & 24.73 & 31.62 & 29.37 & -0.14 & 0.04 & -0.02 & 0.8225 & 0.9341 & 0.9043 & 0.8218 & 0.9337 & 0.9037 & -0.0013 & 0.0000 & -0.0006 \\
\textit{Kitchen} & 244 & 25.62 & 35.67 & 32.30 & 25.63 & 35.61 & 32.28 & -0.30 & 0.14 & -0.02 & 0.7850 & 0.9669 & 0.9404 & 0.7854 & 0.9665 & 0.9400 & -0.0018 & 0.0014 & -0.0004 \\
\textit{Stump} & 109 & 17.63 & 31.94 & 26.52 & 17.64 & 31.93 & 26.52 & -0.06 & 0.07 & 0.00 & 0.5716 & 0.9140 & 0.8011 & 0.5713 & 0.9139 & 0.8008 & -0.0021 & 0.0002 & -0.0004 \\
\textit{Room} & 272 & 28.79 & 36.32 & 32.93 & 28.82 & 36.26 & 32.90 & -0.20 & 0.09 & -0.02 & 0.8633 & 0.9802 & 0.9349 & 0.8634 & 0.9801 & 0.9332 & -0.0058 & 0.0023 & -0.0017 \\
\textit{Playroom} & 196 & 28.32 & 40.32 & 35.07 & 28.36 & 40.28 & 35.03 & -1.38 & 0.26 & -0.04 & 0.9064 & 0.9817 & 0.9548 & 0.9047 & 0.9816 & 0.9542 & -0.0028 & 0.0004 & -0.0006 \\
\textit{drJohnson} & 230 & 29.55 & 42.07 & 36.51 & 29.56 & 41.95 & 36.39 & -5.05 & 0.17 & -0.13 & 0.9221 & 0.9863 & 0.9652 & 0.9156 & 0.9861 & 0.9645 & -0.0129 & 0.0002 & -0.0006 \\
\bottomrule
\end{tabular}
}
\end{table*}

%% file: tables/runtime_microbench_combo.tex
\begin{figure*}[t]
\centering

\begin{minipage}[t]{0.64\textwidth}
\vspace{0pt}
\centering

\parbox[c][1.8em][c]{\linewidth}{%
    \centering
    \small\textbf{(a) CUDA/RT Latency and Speedup}%
}

\vspace{3pt}

\includegraphics[
    width=\linewidth,
    trim=0 0 0 8,
    clip
]{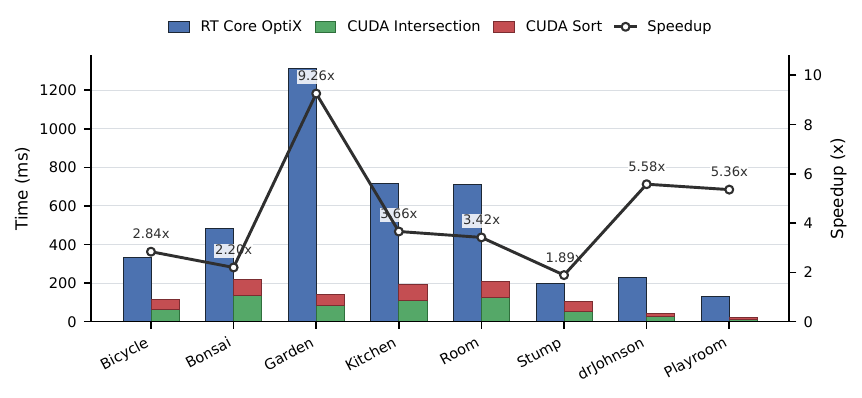}

\end{minipage}
\hfill
\begin{minipage}[t]{0.35\textwidth}
\vspace{0pt}
\centering

\refstepcounter{table}
\label{tab:runtime-microbench}

\parbox[c][1.8em][c]{\linewidth}{%
    \centering
    \small\textbf{Table~\thetable: Simulated RT Kernel Performance}%
}

\vspace{3pt}

\tiny
\setlength{\tabcolsep}{2.2pt}
\renewcommand{\arraystretch}{1.04}

\resizebox{\linewidth}{!}{%
\begin{tabular}{@{}lrrrr@{}}
\toprule

\textbf{Scene}
& \multicolumn{2}{c}{\textbf{Latency}}
& \textbf{Speedup}
& \textbf{DRAM} \\

\cmidrule(lr){2-3}

& \textbf{1-Grid}
& \textbf{27-Grid}
& {}
& {} \\

\midrule

\textit{Bicycle}
& 12,118 & 9,974
& 1.21$\times$
& 21.3\%$\downarrow$ \\

\textit{Bonsai}
& 7,207 & 5,481
& 1.31$\times$
& 31.3\%$\downarrow$ \\

\textit{Garden}
& 19,390 & 17,423
& 1.11$\times$
& 12.1\%$\downarrow$ \\

\textit{Kitchen}
& 7,828 & 6,142
& 1.27$\times$
& 30.9\%$\downarrow$ \\

\textit{Room}
& 4,944 & 4,582
& 1.08$\times$
& 5.9\%$\downarrow$ \\

\textit{Stump}
& 14,993 & 12,500
& 1.20$\times$
& 20.7\%$\downarrow$ \\

\textit{drJohnson}
& 7,104 & 5,919
& 1.20$\times$
& 17.9\%$\downarrow$ \\

\textit{Playroom}
& 5,536 & 5,022
& 1.10$\times$
& 11.1\%$\downarrow$ \\

\bottomrule
\end{tabular}%
}

\vspace{2pt}

\footnotesize
Latency is reported in cycles per ray over proxy center rays sent to the
RT Core traversal engine. Each row is an independent per-scene average over
40 benchmark patches.

\end{minipage}

\caption{
Runtime and traversal microbenchmark summary.
(a) decomposes \HhRay latency into the modeled RT Core component and
measured CUDA components, with per-scene speedup over RaySplats overlaid.
Table~\ref{tab:runtime-microbench} isolates RT-core traversal latency
and DRAM read reduction.
}
\label{fig:runtime-time-results}

\end{figure*}

%% file: tables/drjohnson-outliers.tex
\begin{table}[t]
\centering
\caption{
Diagnostic reruns for the two largest \textit{drJohnson} PSNR outliers.
All values are PSNR in dB; $\Delta$ values are relative to RaySplats.
The no-proxy column reports independent per-ray traversal, while the 2-hop and 3-hop columns use RCPR.
}
\label{tab:drjohnson-outliers}
\scriptsize
\setlength{\tabcolsep}{3pt}
\begin{tabular}{@{}lrrrrrr@{}}
\toprule
\textbf{View}
& \textbf{RaySplats}
& \textbf{2-hop}
& $\Delta$
& \textbf{3-hop}
& $\Delta$
& \textbf{No proxy} \\
\midrule
\texttt{img\_6486} & 39.86 & 34.81 & -5.05 & 35.05 & -4.82 & 34.95 \\
\texttt{img\_6541} & 37.60 & 35.81 & -1.79 & 34.68 & -2.92 & 35.45 \\
\bottomrule
\end{tabular}
\end{table}

%% file: tables/sensitivity.tex
\begin{table}[t]
\caption{Hop-distance PSNR sensitivity.}
\label{tab:hop-distribution}
\centering
\footnotesize

\setlength{\tabcolsep}{4pt}

\resizebox{\columnwidth}{!}{%
\begin{tabular}{c cccc}
\toprule

\textbf{Scene}
& \textbf{Baseline}
& \textbf{0-hop}
& \textbf{1-hop}
& \textbf{2-hop} \\

\midrule

\textit{Bicycle}
 & 31.10
 & 28.07 (-3.03)
 & 29.96 (-1.14)
 & 31.05 (-0.05) \\

\textit{Bonsai}
 & 35.81
 & 35.51 (-0.30)
 & 35.75 (-0.06)
 & 35.76 (-0.05) \\

\textit{Garden}
 & 31.66
 & 26.11 (-5.55)
 & 30.38 (-1.28)
 & 31.64 (-0.02) \\

\textit{Kitchen}
 & 35.71
 & 35.22 (-0.49)
 & 35.59 (-0.12)
 & 35.64 (-0.07) \\

\textit{Room}
 & 36.36
 & 35.99 (-0.37)
 & 36.24 (-0.12)
 & 36.29 (-0.07) \\

\textit{Stump}
 & 31.97
 & 31.69 (-0.28)
 & 31.89 (-0.08)
 & 31.96 (-0.01) \\

\textit{drJohnson}
 & 42.11
 & 38.40 (-3.71)
 & 41.26 (-0.85)
 & 41.98 (-0.13) \\

\textit{Playroom}
 & 40.36
 & 30.17 (-10.19)
 & 33.42 (-6.94)
 & 40.32 (-0.04) \\

\midrule

\textbf{Mean Diff (dB)}
& -
& \textbf{-2.99}
& \textbf{-1.32}
& \textbf{-0.06} \\

\bottomrule
\end{tabular}
}

\vspace{2pt}
\scriptsize
Baseline is the C++ full per-ray traversal renderer without RCPR, not RaySplats.
All hop settings use the same selected view and ground truth per scene.
Parentheses denote dB differences from baseline; the final row reports their arithmetic mean.

\end{table}

%% file: tables/fisheye_object.tex
\begin{figure*}[t]
\centering

\setlength{\tabcolsep}{2pt}

\scriptsize

\resizebox{\textwidth}{!}{%
\begin{tabular}{lcccc}

\toprule

& \textbf{Perspective}
& \textbf{Perspective + Mesh}
& \textbf{Fisheye}
& \textbf{Fisheye + Mesh} \\

\midrule

&
\includegraphics[width=0.225\textwidth]{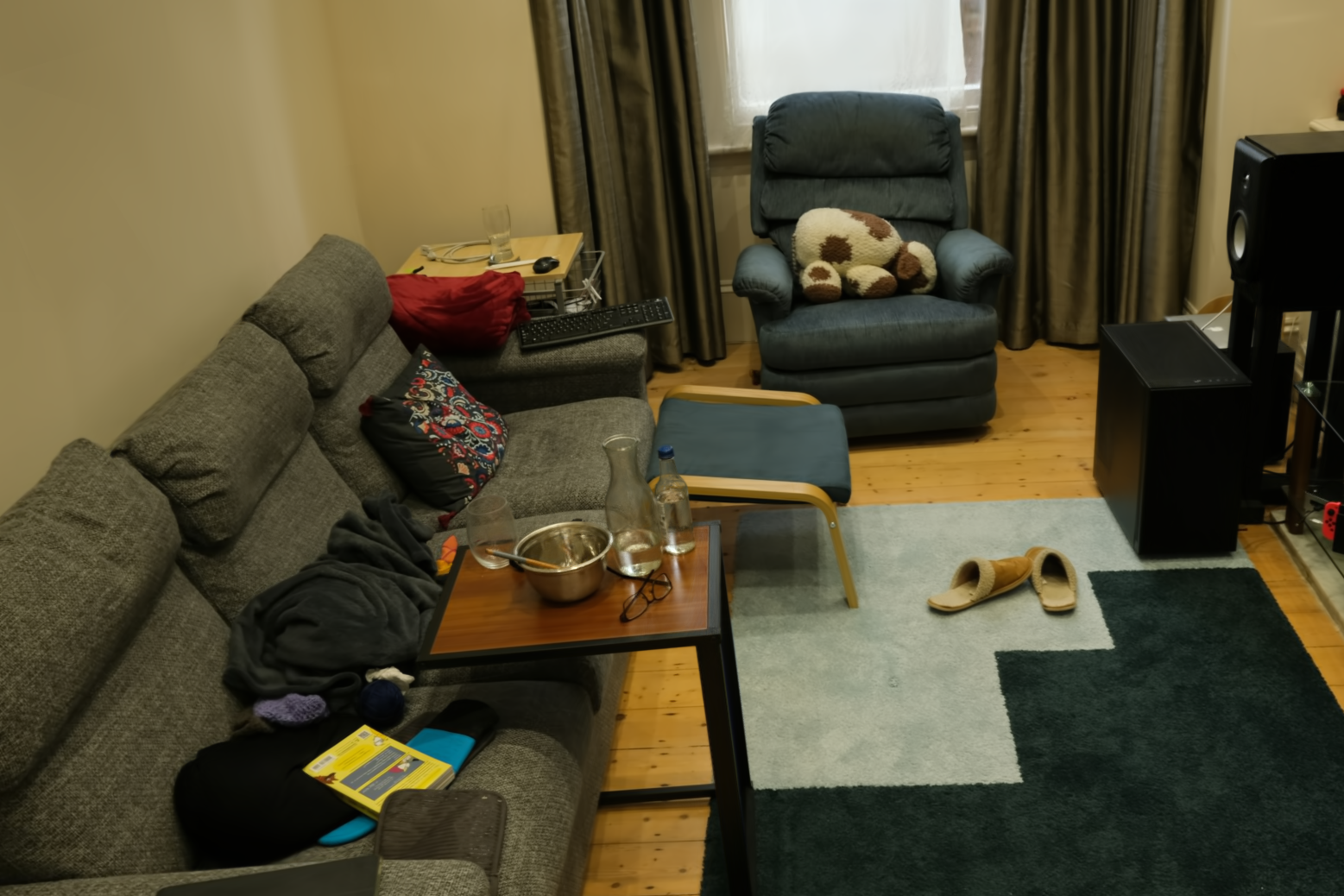}
&
\includegraphics[width=0.225\textwidth]{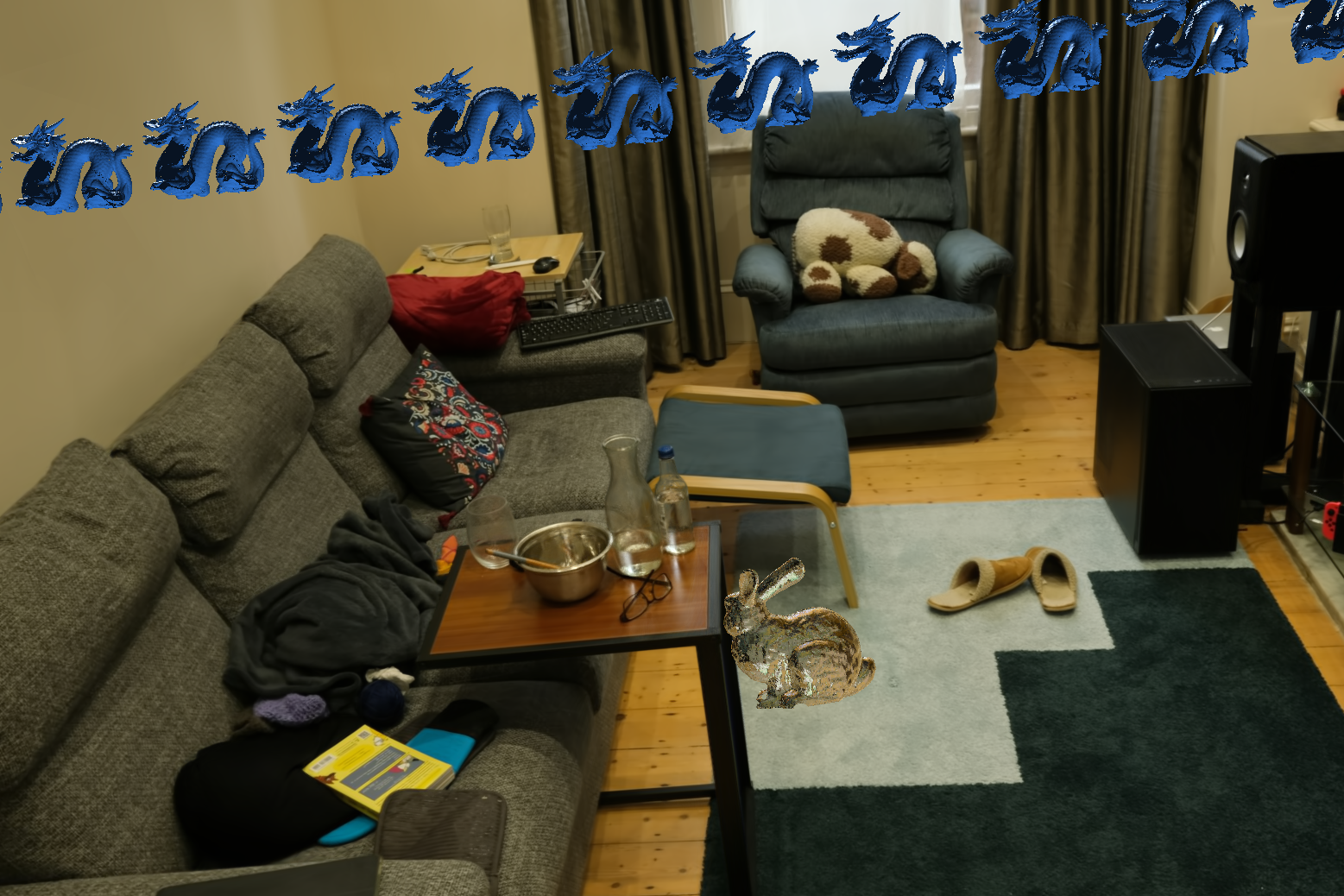}
&
\includegraphics[width=0.225\textwidth]{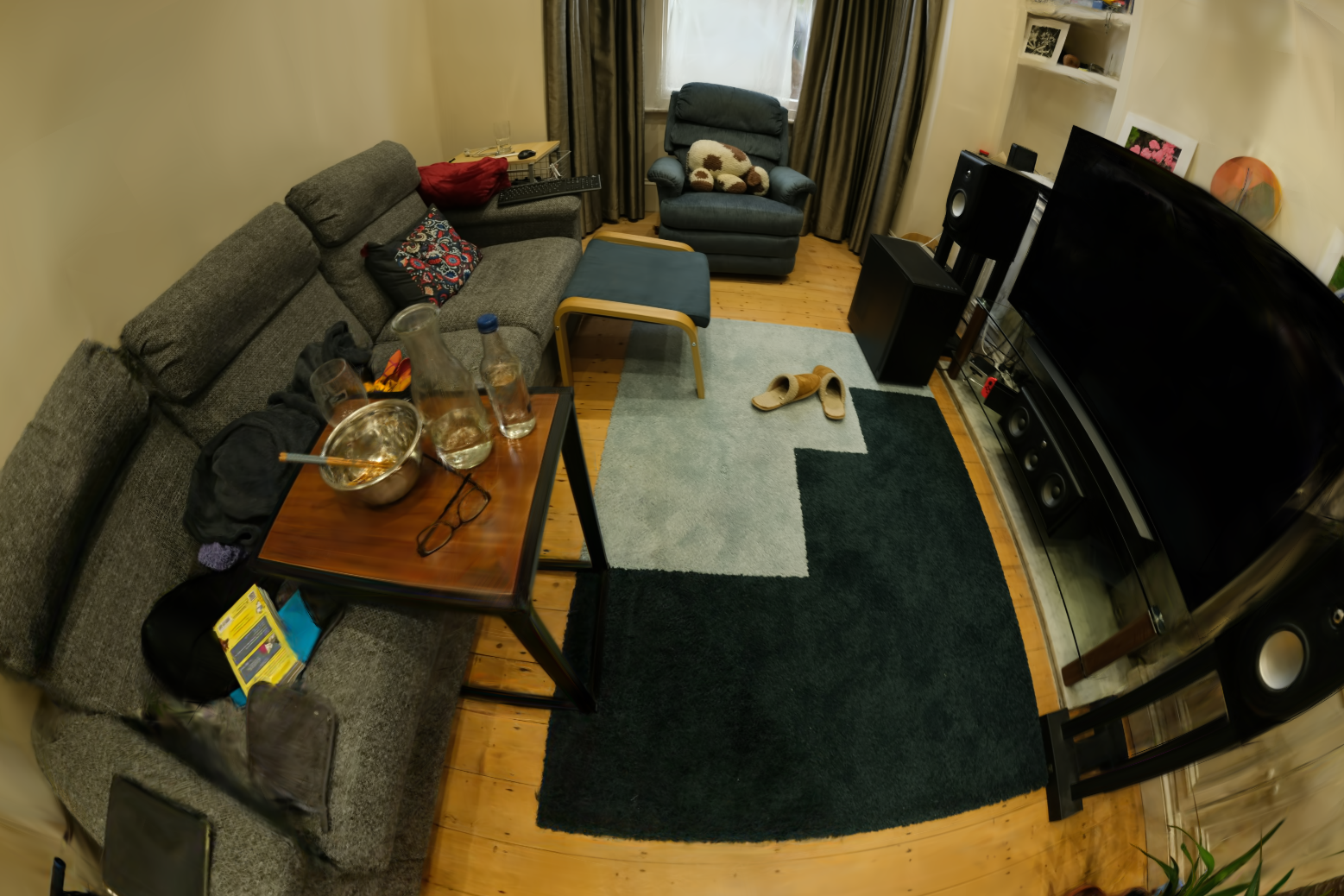}
&
\includegraphics[width=0.225\textwidth]{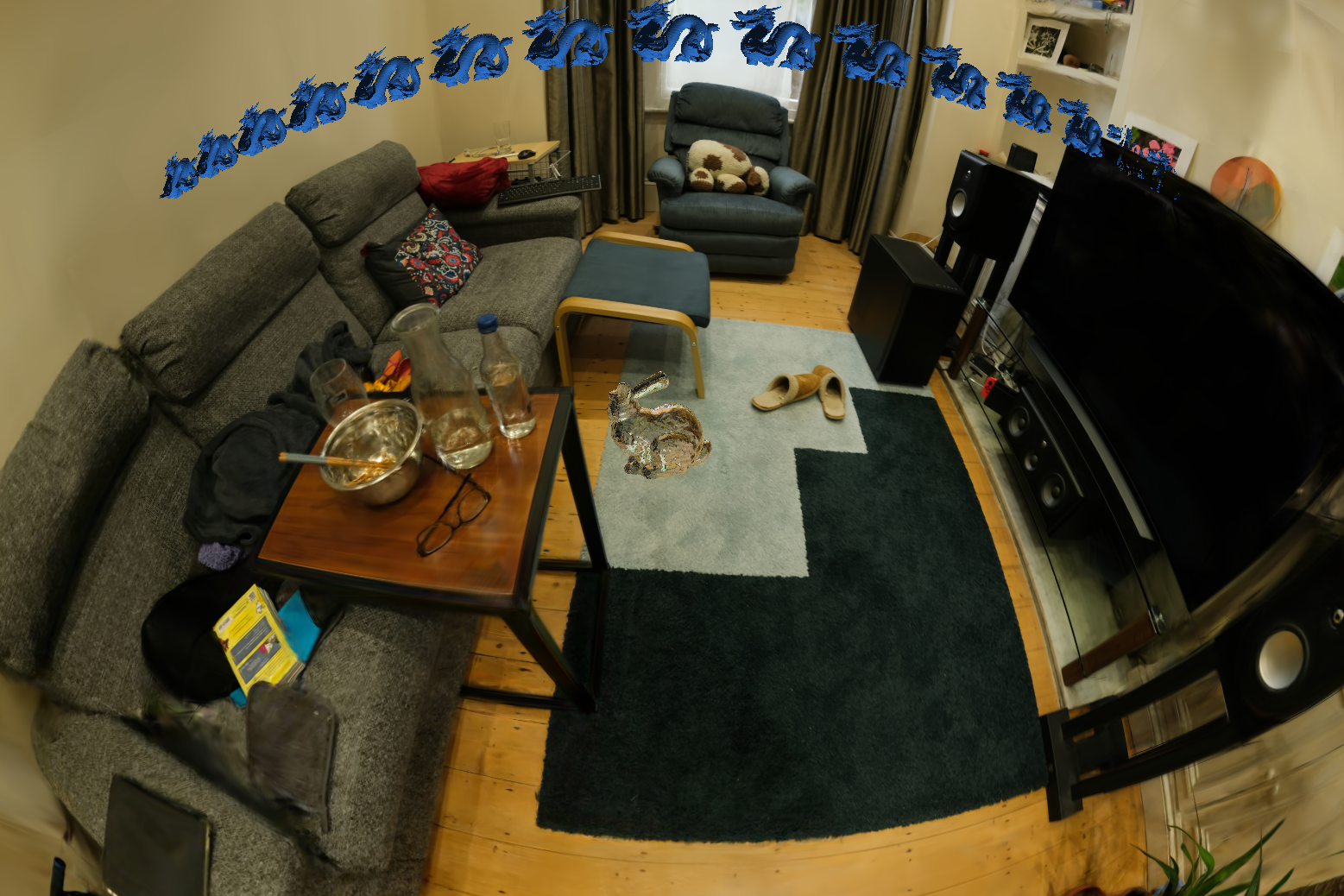}
\\

\midrule

\textbf{ray--AABB}
& 9,156,983,251
& 9,344,892,164
& 7,744,344,921
& 7,842,369,262
\\

\textbf{ray--Gaussian}
& 1,873,882,463
& 1,898,858,274
& 1,741,258,064
& 1,754,039,606
\\

\textbf{ray--triangle}
& -
& 1,380,169
& -
& 731,771
\\

\bottomrule
\end{tabular}
}

\normalsize

\caption{
Mixed rendering workloads and corresponding ray--primitive intersection statistics.
The first row shows rendered images for the \textit{Room} scene under different rendering modes at the evaluation resolution, using the default 2-hop and $3^3$ grid setting.
The following rows report the number of ray--primitive tests,
including ray--AABB, ray--Gaussian, and ray--triangle tests.
}

\label{fig:uie-mixed-rendering-stats}

\end{figure*}

%% file: sections/07-RelatedWorks.tex
\section{Related Work}

\paragraph{Neural Rendering Accelerator}
NeRF accelerators target repeated point sampling and neural-field evaluation. Prior designs optimize multiresolution encoding and MLP inference~\cite{mubarikHardwareAccelerationNeural2023}, fused scene learning~\cite{songCambriconR2023}, and sampling through co-design~\cite{liRTNeRFRealTimeOnDevice2022,fuGenNeRFEfficientGeneralizable2023}. \HhRay instead accelerates explicit Gaussian traversal and intersection.

\paragraph{3D Gaussian Splatting Accelerator}
GauSPU~\cite{wuGauSPU3DGaussian2024}, GauRast~\cite{liGauRastEnhancingGPUTriangle2025}, Splatonic~\cite{huangSplatonicArchitectureSupport2026}, ORANGE~\cite{liORANGEExploringOckhams2026}, and Cambricon-GS~\cite{wenCambriconGSAccelerator2026} accelerate SLAM, rasterization, sparse pixels, blending, or training. Unlike \HhRay, these systems retain splatting-style projection rather than ray-traced visibility.

\paragraph{Ray Tracing Accelerator}
NVIDIA RT Cores~\cite{nvidiaTuringRT2018,nvidiaAdaRT2022} and AMD RDNA units~\cite{amdRDNA2025} accelerate BVH traversal and triangle tests. GRTX~\cite{GRTX} adds Gaussian BVHs and single-ray traversal checkpointing.\footnote{GRTX has no public implementation at the time of writing, so we cannot directly compare against it.} GRTX stores breakpoints along one ray, whereas RCPR lets independent neighboring rays reuse a nearby ray's traversal path; these optimizations are orthogonal. \HhRay further adds LB-SBVH for BVH partitioning and UIE for Gaussian--triangle datapaths.
These systems therefore optimize different levels of the pipeline: prior accelerators primarily change representation or primitive execution, whereas \HhRay co-designs scene partitioning, inter-ray reuse, and mixed-primitive support around direct ray--Gaussian traversal.